\documentclass[rmp,aps,nofootinbib,twocolumn]{revtex4}

\usepackage{graphicx}
\usepackage{amsmath,amssymb,bbold,bm}
\usepackage{epstopdf}
\usepackage[varg]{txfonts}
\usepackage{dsfont}

\makeatletter
\def\underbracket{%
\@ifnextchar[{\@underbracket}{\@underbracket [\@bracketheight]}%
}
\def\@underbracket[#1]{%
\@ifnextchar[{\@under@bracket[#1]}{\@under@bracket[#1][0.4em]}%
}
\def\@under@bracket[#1][#2]#3{
\mathop{\vtop{\m@th \ialign {##\crcr $\hfil \displaystyle {#3}\hfil $%
\crcr \noalign {\kern 3\p@ \nointerlineskip }\upbracketfill {#1}{#2}
   \crcr \noalign {\kern 3\p@ }}}}\limits}
\def\upbracketfill#1#2{$\m@th \setbox \z@ \hbox {$\braceld$}
    \edef\@bracketheight{\the\ht\z@}\bracketend{#1}{#2}
    \leaders \vrule \@height #1 \@depth \z@ \hfill
    \leaders \vrule \@height #1 \@depth \z@ \hfill \bracketend {#1}{#2}$}
\def\bracketend#1#2{\vrule height #2 width #1\relax}
\makeatother

\def\inbar{\,\vrule height1.5ex width.4pt depth0pt}
\def\IR{\relax{\rm I\kern-.18em R}}
\def\IC{\relax\hbox{$\inbar\kern-.3em{\rm C}$}}

\def\nuc#1#2{${}^{#1}$#2}
\def\mee{$\langle m_{\beta\beta} \rangle$}

\def\BBz{$\beta\beta(0\nu)$}

\def\BBt{$\beta\beta(2\nu)$}
\def\BB{$\beta\beta$}
\def\Gz{$G^{0\nu}$}
\def\Mz{$M_{0\nu}$}

\def\Tz{$T^{0\nu}_{1/2}$}

\def\MJ{{\sc Majorana}}             
\def\be{\begin{equation}}
\def\ee{\end{equation}}
\def\today{\space\number\day\space\ifcase\month\or January\or February\or
    March\or April\or May\or June\or July\or August\or September\or October\or
    November\or December\fi\space\number\year}
\newcommand{\ket}[1]{\left| #1 \right\rangle}

\newcommand{\bq}{{\bm q}}
\newcommand{\br}{{\bm r}}
\newcommand{\bR}{{\bm R}}
\newcommand{\bB}{{\bm B}}

\newcommand{\bS}{{\bm S}}
\newcommand{\bx}{{\bm x}}

\newcommand{\bsig}{{\bm \sigma}}

\newcommand{\cT}{{\cal T}}

\begin{document}

\title{{\em Colloquium:} Majorana Fermions in Nuclear, Particle and Solid-state Physics}

\author{Steven R. Elliott}
\email{elliotts@lanl.gov}
\affiliation{Los Alamos National Laboratory, Los Alamos, NM, USA, 87545}
\author{Marcel Franz}
\email{franz@physics.ubc.ca}
\affiliation{Department of Physics and Astronomy, University of British Columbia,
Vancouver, BC, Canada, V6T 1Z1}

\begin{abstract}
Ettore Majorana (1906-1938) disappeared while traveling by ship from Palermo to Naples in 1938. His fate has never been fully resolved and several articles have been written that explore the mystery itself.  His demise intrigues us still today because of his seminal work, published the previous year, that established symmetric solutions to the Dirac equation that describe a fermionic particle that is its own anti-particle. This work has long had a significant impact in neutrino physics, where this fundamental question regarding the particle remains unanswered. But the formalism he developed has found many uses as there are now a number of candidate spin-$\frac{1}{2}$ neutral particles that may be truly neutral with no quantum number to distinguish them from their anti-particles. If such particles exist, they will influence many areas of nuclear and particle physics. Most notably the process of neutrinoless double beta decay can only exist if neutrinos are massive Majorana particles. Hence, many efforts to search for this process are underway. Majorana's influence doesn't stop with particle physics, however, even though that was his original consideration. The equations he derived also arise in solid state physics where they describe electronic states in materials with superconducting order. Of special interest here is the class of solutions of the Majorana equation in one and two spatial dimensions at exactly zero energy. These Majorana zero modes are endowed with some remarkable physical properties that may lead to advances in quantum computing and, in fact, there is evidence that they have been experimentally observed. This review first summarizes the basics of Majorana's theory and its implications. It then provides an overview of the rich experimental programs trying to find a fermion that is its own anti-particle in nuclear, particle, and solid state physics.
\end{abstract}


\maketitle
\tableofcontents

\section{Introduction}
\label{Sec:intro}
In the late 1920's Schr\"{o}dinger published  his non-relativistic wave equation~\cite{Schrodinger1926} describing the quantum behavior of fundamental particles. Soon thereafter Paul Dirac developed the wave equation~\cite{Dirac1928} that bears his name describing the behavior of relativistic particles. About a decade later, Ettore Majorana recognized the importance of a specific representation of the Dirac equation~\cite{Majorana1937}. There are numerous examples of elementary particles that are described by the Dirac equation, but as yet, none have been found that obey that of Majorana. Discovering particles or quasi-particles that are governed by Majorana's formalism would have significant implications for science. Depending on the type of particle or quasi-particle found, subfields ranging from cosmology to particle physics to solid state physics would be affected. The remarkable achievement of the {\em Majorana equation} and its potential consequences motivate this colloquium.

\subsection{The Particle Physics View}
\label{Sec:PartPhysView}
Schr\"{o}dinger's equation was developed using the classical equation relating energy ($E$) to momentum (${\bm p}$) for a particle of mass $m$
\begin{equation}
E = {\bm p}^2/2m,
\end{equation}
and substituting the corresponding differential operators ($\hbar$=c=1):
\begin{equation}
E \rightarrow i\frac{\partial}{\partial t},	\ \ \ \
\bm{p} \rightarrow -i\nabla
\label{eqn:operators}
\end{equation}
to find the resulting wave equation,
\begin{equation}
i\dot{\Psi} = -\frac{1}{2m}\nabla^2\Psi
\end{equation}
where $\Psi$ is the wavefunction. Relativistically, the energy-momentum relation is
\begin{equation}
E^2 = p^2 + m^2
\label{eqn:EnMo}
\end{equation}
and a straight-forward substitution of the operators in Eq.~\ref{eqn:operators} leads to the Klein-Gordon equation which has a double differentiation with respect to time. This feature of the Klein-Gordon equation, which differs from the Schr\"{o}dinger equation, means that the probability of the value of a dynamic variable cannot be predicted at a future time when $\Psi$ is provided at a given earlier time. Dirac, wanting to avoid this feature, succeeded in writing an equation that was linear in $\dot{\Psi}$, removing this difficulty. Dirac wrote his equation as
\begin{equation}
\label{eqn:Dirac}
i\dot{\Psi} = H_{\rm Dirac}\Psi = (\bm{\alpha}\cdot\bm{p} + \beta m)\Psi.
\end{equation}
The $\alpha$'s and $\beta$ of Eq.~(\ref{eqn:Dirac}) do not commute and hence they cannot be simple numbers, but Dirac was able to find a set of $4\times4$ matrices that met the requirements for his equation. The form of those matrices are such that operating with $H_{Dirac}$ twice will result in Eq.~\ref{eqn:EnMo}. The solutions of Eq.~(\ref{eqn:Dirac}) are 4-component vectors that describe a spin-$\frac{1}{2}$ particle that is distinct from its anti-particle. Such anti-particles, including the electron's anti-particle partner the positron and neutron's anti-particle partner the anti-neutron, were found soon after Dirac's publication, and his work has been considered a prediction of their existence.

A more modern description expresses the Dirac equation so that it is manifestly Lorentz covariant. One can write Eq.~(\ref{eqn:Dirac}) as such by multiplying the equation by the matrix $\beta$ and defining the Dirac representation matrices $\gamma^{\mu} = (\beta;\beta\alpha_i)$. The equation is then
\begin{equation}
\label{eqn:DiracCov}
(i\gamma^{\mu}\partial_{\mu} - m)\Psi = (\gamma^{\mu}p_{\mu} - m)\Psi = 0.
\end{equation}
Note that $p^{\mu} = (i\partial_t,\bm{p})$, $p_{\mu} = (i\partial_t,-\bm{p})$ and $\partial_{\mu}$ is shorthand for ${\partial}/{\partial x^{\mu}}$. The explicit form of the Dirac matrices can be written in terms of the $2\times 2$ Pauli matrices ($\sigma^i$),
\begin{equation}
\sigma^1 = \left( \begin{array}{cc} 0 & 1 \\ 1 & 0 \end{array} \right), \  
\sigma^2 = \left( \begin{array}{cc} 0 & -i \\ i & 0 \end{array} \right), \ 
\sigma^3 = \left( \begin{array}{cc} 1 & 0 \\ 0 & -1 \end{array} \right),
\end{equation}
and $\sigma^0$ is the identity matrix. Denoting $\sigma^{\mu} = (\sigma^0,-\sigma^i)$ and $\hat{\sigma}^{\mu} = (\sigma^0,\sigma^i)$, the Dirac matrices can be written, in the Weyl representation, as;
\begin{equation}
\label{eqn:DiracMatrices}
\gamma^{\mu} =  \left( \begin{array}{cc}0 & \sigma^{\mu} \\ \hat{\sigma}^{\mu} & 0\end{array} \right)
\end{equation}
The choice of the matrices in Eq.~(\ref{eqn:DiracMatrices}), however, is not unique. Majorana's insight was that for a particular choice of the $\alpha$'s and $\beta$, Eq.~(\ref{eqn:Dirac}) is real with a real solution. The corresponding Dirac matrices in Majorana representation read
\begin{eqnarray}\label{eqn:MajoranaMatrices}
\tilde{\gamma}^0 = i\begin{pmatrix}0 & -\sigma^1 \\ \sigma^1 & 0 \end{pmatrix}, \ \ \ 
\tilde{\gamma}^1 = i\begin{pmatrix}0 & \sigma^0 \\ \sigma^0 & 0 \end{pmatrix},  \nonumber \\
\tilde{\gamma}^2 = i\begin{pmatrix} \sigma^0 & 0 \\ 0 & -\sigma^0 \end{pmatrix}, \ \ \ 
\tilde{\gamma}^3 = \begin{pmatrix}0 & \sigma^2 \\ -\sigma^2 & 0 \end{pmatrix}. 
\end{eqnarray}

The Dirac equation is actually four coupled equations for 4 spinor components. For the Majorana equation, the condition of reality reduces this to two independent systems, each with two coupled equations. The solution to one of these systems then describes a truly neutral particle, still spin $\frac{1}{2}$, but with no distinction between particle and anti-particle. Such a particle, if found, would be termed a Majorana fermion.\footnote{Majorana's paper was published a year before his disappearance from a transport ship between Palermo and Naples. The story behind his life and theories of his demise are interesting in their own right and are summarized by Holstein~\cite{Holstein2009} and references therein.} Eddington~\cite{Eddington1928} noted that one could derive symmetrical equations from the Dirac equation, but it was Majorana who noted the particle/anti-particle correspondence and its importance for neutrinos.

The four components of a Dirac wave function describe a particle and anti-particle pair, each with spin $\frac{1}{2}$. The equivalent in the Majorana picture are two particles, each of spin $\frac{1}{2}$. In some sense one can consider a Dirac fermion as a special case of a Majorana fermion pair. Two, mass-degenerate Majorana fermions with opposite CP parity\footnote{CP refers to combined operators of charge conjugation and spatial inversion symmetry. When the operator $\cal CP$ acts upon a wave function, it changes the sign of all the position coordinates and changes the particle to anti-particle. For a Majorana fermion it is somewhat more subtle as ${\cal C}^2\ket{\nu} \equiv \ket{\nu}$, and therefore ${\cal CP}\ket{\nu} = \pm i\ket{\nu}$~\cite{Carruthers1971,Kayser1983}} would be indistinguishable from a Dirac fermion with that same mass.

\subsection{The Solid State Physics View}
\label{Sec:IntroSolidState}
In solid state physics the only fermionic particles that matter for
all practical purposes are electrons. Electrons are, of course, Dirac
fermions. However, as it turns out, Majorana fermions can occur in
certain solids as {\em emergent quasiparticles} which can be thought
of as collective excitations of the quantum many-body state describing the
interacting electron system.      

Condensed matter physics is replete with examples of
emergence  \cite{Anderson1972}. Emergent particles in solids range from those very well
established (phonons, magnons, plasmons, polarons) through some that are more elusive
(triplons, composite fermions) to truly exotic and speculative (spinons, holons,
chargons, visons, etc.). At the
conventional end, phonons for example represent quanta of the lattice vibrations,
and form an essential ingredient in the description of the
low-temperature thermodynamic and transport properties of all
solids \cite{Kittel1987}. Magnons -- quanta of spin fluctuations -- are similarly essential
in the description of magnetic solids, as are polarons for ionic
insulators and semiconductors. In this sense, emergent
particles are as real as the elementary particles in nuclear and
high-energy physics. Observation of an emergent Majorana fermion
in a solid state system would be as exciting, and perhaps even more so, as establishing
that e.g.\ the neutrino is a Majorana fermion. As we shall see in Sections \ref{Sec:Theory} and \ref{Sec:SolidState} quasiparticle excitations in superconductors indeed behave in all respects as Majorana fermions.

To understand how Majorana fermions emerge in a system comprised of
many electrons we must first briefly review the structure of such
electronic many-body states. These are efficiently described using the 
formalism of second quantization which is uniquely suited to handle 
systems with a very large numbers of identical
particles in condensed matter physics. In this formalism electrons are represented by a set of
creation and annihilation operators where $c_j^\dagger$ creates an electron
with quantum numbers denoted by index $j$ while
$c_j$ annihilates it. In a typical situation $j$ includes the position
degree of freedom as well as the orbital and spin quantum numbers.
As appropriate for identical fermions these operators obey the
canonical anticommutation relations
\begin{equation}\label{can0}
\{c_i^\dagger,c_j^\dagger\}=\{c_i,c_j\}=0, \ \ \ \ 
\{c_i^\dagger,c_j\}=\delta_{ij}.
\end{equation}
A Hamiltonian describing electrons in an arbitrary solid can be expressed in terms
these operators; the electron kinetic energy will be
represented by terms bi-linear in $c$'s while interactions will
contain quartic terms.

 Without any loss of generality, one can perform
a canonical transformation of the Hamiltonian (and any other operator
of interest) to the ``Majorana basis'', 
\begin{equation}\label{can1}
c_j={1\over 2}(\gamma_{j1}+i\gamma_{j2}), \ \ \ c_j^\dagger={1\over 2}(\gamma_{j1}-i\gamma_{j2}),
\end{equation} 
where the new operators $\gamma_{j\alpha}$, which can be loosely
thought of as the real and the imaginary part of the electron operator, satisfy the following
algebra
\begin{equation}\label{can2}
\{\gamma_{i\alpha},\gamma_{j\beta}\}=2\delta_{ij}\delta_{\alpha\beta},
\ \ \ \gamma_{i\alpha}^\dagger=\gamma_{i\alpha}.
\end{equation}
The last relation informs us that a particle created by the
$\gamma$-operator is identical to its antiparticle: creating and
destroying such a particle has the same effect on the state of the
system. This is a Majorana fermion.

The above discussion shows that {\em any} system of electrons can be
formally recast in terms of Majorana fermion operators through the canonical
transformation (\ref{can1},\ref{can2}). In most cases, however, such a
transformation brings no benefit and merely complicates
things. Physically this is because in most cases the two Majoranas
comprising a given electron are intertwined in space and it thus makes
little sense to describe them as separate entities. There is, however, a special
class of systems, called topological superconductors, in which two
Majorana fermions comprising a single electron become spatially
separated. In this case a description through the Majorana basis
becomes essential as no other basis can accurately account for the true
physical degrees of freedom in the system. It is precisely this class
of systems that we will discuss in great detail in Sec.\
\ref{Sec:SolidState} since they exhibit, in very real and quantifiable sense, independent Majorana particles.  

There are a number of conditions that must be satisfied for a system of
electrons to exhibit unpaired  Majorana fermions. One key condition that can be
easily understood follows immediately from inverting the
transformation specified in Eq.\ (\ref{can1}) to obtain
\begin{equation}\label{can3}
\gamma_{j1}=c_j^\dagger+c_j, \ \ \ \ \gamma_{j2}=i(c_j^\dagger-c_j).
\end{equation}
These relations suggest that isolated Majorana fermions can be found in
systems with superconducting order. This is because coherent
superpositions of electron and hole degrees of freedom, indicated in
Eqs.\ (\ref{can3}), are known to naturally occur only in the theory of superconductivity,
originally due to Bardeen, Cooper and Schrieffer \cite{BCS1957}. Also,
an operator defined in Eq.\ (\ref{can3}) can only act nontrivially on
a ground state with uncertain total number of particles; such a ground state is characteristic
of superconducting systems.
It therefore follows that one needs a superconductor to observe Majorana
fermions in the solid-state context, or an interacting many-body
system whose effective description is that of a superconductor \cite{Read2000}.

Whether or not a neutrino (or another elementary particle) is a
Majorana fermion remains ultimately an experimental question: theory
allows such a possibility but experiment will have the final word. In
solid state physics the situation is somewhat different. The relevant
theory, based on the band theory of solids and the BCS theory of
superconductivity, both exceptionally well understood and tested,
unambiguously predicts that Majorana fermions should exist in superconductors.
As we shall discuss in more detail below, there is in fact good experimental evidence that quasiparticle excitations that occur in superconductors at non-zero energies are described by the variant of the Majorana equation. A point that remains to be experimentally settled is the existence of {\em Majorana zero modes}. These constitute a special case of Majorana particles occurring at exactly zero energy. They are thought to be endowed with some very unusual physical properties that have no analog in high-energy physics.

 There is now no doubt that
Majorana zero modes should emerge in a class of systems called ``topological superconductors''. The uncertainty has to do with the
question of whether or not the experiments have achieved conditions necessary for the occurrence of topological superconductivity. Further uncertainty arises from the difficulties related to the unambiguous detection of the Majorana zero modes in solid-state devices, where their signatures can be masked by the effect of disorder or mimicked by other unrelated effects. At the time of this writing a consensus is building up that Majorana zero modes have been observed in recent experiments on semiconductor quantum wires proximity-coupled to a superconductor \cite{Mourik2012}. Experiments on other systems that have been proposed to host Majorana particles are being actively pursued.  In Sec.\
\ref{Sec:SolidState} we shall review the underlying theory and describe the relevant experimental work as well as discuss in more detail various uncertainties that still exist.

\subsection{Significance and Potential Applications}
\label{Sec:IntroSignif}
The broad interest in the physics community regarding the existence of Majorana states, whether they be true particles or an emergent quantum state in condensed matter systems, arises because of the significance of such states and their potential applications. Here we mention two important potential outcomes of the existence of Majorana fermions: leptogenesis and quantum computing.

Observationally, the Universe is composed of matter with no, or little, anti-matter. This fact is necessary for us to exist in order to even ponder it. There must have been an initial imbalance between the two, or otherwise all matter would have annihilated with anti-matter leaving nothing to form the galaxies, solar systems, planets and, of course, us. The Big Bang, however, made an equal number of matter and anti-matter particles at the Universe's beginning. The final asymmetry between matter and anti-matter arose dynamically after the Big Bang. If neutrinos are Majorana particles, the theory of leptogenesis might explain this asymmetry. We discuss this possibility in Sec.~\ref{sec:leptogenesis}.

In a solid state system, Majorana fermions naturally occur in superconductors  and can appear as isolated and unpaired zero modes in topological superconductors and certain fractional quantum Hall systems. The state vector that describes such Majorana zero modes that are spatially separated can be used to encode quantum information. Since they are spatially separated, the information in each quantum bit is stored nonlocally leading to long decoherence times, a necessary feature for robust quantum computing. We discuss this possibility in Sec.~\ref{sec:NonAbelian}.

\section{The Majorana Equation and its Consequences}
\label{Sec:Theory}

The Dirac equation, its symmetries, solutions and physical
consequences are discussed in standard textbooks
\cite{Peskin1995}. The Majorana solution to the Dirac equation has
likewise been reviewed in a great detail \cite{Chamon2010,Pal2011}. In
this Section we outline the key theoretical ideas behind the concept
of Majorana fermions emphasizing similarities and differences
 between the elementary  particle and condensed matter physics.

\subsection{The Majorana-Dirac Equations}
The solution of the free-particle Dirac equation (\ref{eqn:DiracCov}) is a four-component bi-spinor,
\begin{equation}
\Psi = \left( \begin{array}{c}
\psi_1 \\
\psi_2 \\
\psi_3 \\
\psi_4 \end{array} \right).
\end{equation}
If the spin vector of a particle points in the same direction as its momentum, it is referred to as right-handed. Defining $\psi_R$ and $\psi_L$ as the two component spinors that are the right-handed and left-handed projections of $\psi$ (chiral projections), 
\begin{equation}
\Psi \equiv \left( \begin{array}{c} \psi_R \\ \psi_L \\\end{array} \right),
\end{equation}
we use the Weyl representation (\ref{eqn:DiracMatrices}) to write the Dirac equation (\ref{eqn:DiracCov}) as
\begin{eqnarray} \label{Eqn:DirEqn}
(i\partial_t - \bm{p}\cdot \bm{\sigma})\psi_R - m_D\psi_L & = & 0 \\	\nonumber
(i\partial_t + \bm{p}\cdot \bm{\sigma})\psi_L - m_D\psi_R & = & 0,
\end{eqnarray}
where we have suggestively added a subscript $D$ for Dirac to the mass ($m_D$).  

Dirac equation (\ref{Eqn:DirEqn}) has many symmetries that have been discussed at length in the literature. Of these, two will be most important for our discussion of Majorana particles. The {\em global gauge} symmetry, expressed as
\begin{equation} \label{gauge}
\Psi(x)\to e^{i\theta}\Psi(x)
\end{equation}
 with $\theta$ a real constant, guarantees that one can couple the Dirac fermion to the electromagnetic field and thus describe charged particles. This is accomplished by the minimal substitution $p_\mu\to p_\mu-eA_\mu$ where $A_\mu$ represents the electromagnetic gauge potential. 

The {\em charge conjugation} symmetry $\cal C$ is best understood by examining the stationary solutions of Eq. (\ref{Eqn:DirEqn}). If we separate the time dependence as $\Psi(x)=e^{-iEt}\Phi(\bx)$, the Dirac equation for $\Phi(\bx)$ retains the form displayed in  Eq. (\ref{Eqn:DirEqn}) with $i\partial_t$ replaced by energy $E$. This stationary Dirac equation has the following property: for each solution  $\Phi(\bx)$ with energy $E$ there exists a solution 
\begin{equation} \label{cc}
\Phi^c(\bx)={\cal C}\Phi(\bx)\equiv C\Phi^*(\bx),
\end{equation}
with energy $-E$. Here $C$ is the charge conjugation matrix and * denotes complex conjugation. In the Weyl representation (\ref{eqn:DiracMatrices}) of the Dirac matrices $C=i\gamma^2$ while in the Majorana representation (\ref{eqn:MajoranaMatrices}) $C=\ensuremath{\mathds{1}}$. For charged particles it is easy to show that the $\cal C$ operation reverses the sign of the particle charge $e$, hence its name. 

Traditionally, the positive energy solutions of the Dirac equation are taken to describe particles (e.g.\ electrons) while the negative energy solutions the corresponding antiparticles (e.g.\ positrons).
In order to describe a particle that is indistinguishable from its antiparticle we demand, following Majorana, that 
\begin{equation} \label{ccm}
\Psi^c(x)=\Psi(x),
\end{equation}
i.e.\ that the particle wavefunction and its charge conjugate partner are the same. Importantly, we note that this condition can only be met when imposed on  the time-dependent wavefunction $\Psi(x)$. For $E\neq 0$ stationary solutions $\Phi(\bx)$ and  $\Phi^c(\bx)$ belong to different energy eigenvalues and are thus necessarily orthogonal. In the special case $E=0$ the Majorana condition (\ref{ccm}) can be satisfied even by a stationary state $\Phi(\bx)$ leading to the concept of the Majorana zero mode which has been of great interest in condensed matter physics and will be discussed at length in Sec.\ IV.

Putting all this together, the corresponding Majorana version of the Dirac equations (\ref{Eqn:DirEqn}) are
\begin{eqnarray}
\label{Eqn:MajEqn}
(i\partial_t - \bm{p}\cdot \bm{\sigma})\psi_R - im_R\sigma^2\psi_R^* & = & 0 \\	\nonumber
(i\partial_t + \bm{p}\cdot \bm{\sigma})\psi_L - im_L\sigma^2\psi_L^* & = & 0.
\end{eqnarray}
Here we have explicitly indicated that the masses are not required to be equal since the Majorana equation decouples. This is not so for the Dirac equation. One should recognize that when the mass is zero, the two equations are equivalent. Although not essential for our discussion, Majorana states are often described in the literature as eigenstates of $\cal CP$.  We note that if a pair of Majorana particles have equal masses but opposite CP parity, that pair would be indistinguishable from a Dirac particle. The two fields $\psi_L$ and $\psi_R$ are then both eigenstates of $\cal CP$ with opposite eigenvalues or CP parities. 

It is also important to notice that unlike the Dirac equation (\ref{Eqn:DirEqn}), the Majorana equation (\ref{Eqn:MajEqn}) is not invariant under the global gauge transformation Eq.\ (\ref{gauge}). Majorana particles cannot be coupled to the electromagnetic field and are thus necessarily charge neutral. Indeed they can carry no quantum numbers that distinguish particles from antiparticles. 

The Majorana fields can be expressed in 4-component form as
\begin{equation}
\label{Eqn:PsiLR}
\Psi_L = \left( \begin{array}{c} -i\sigma^2\psi_L^* \\ \psi_L \end{array}\right), 
\Psi_R = \left( \begin{array}{c} \psi_R \\ i\sigma^2\psi_R^* \end{array}\right).
\end{equation}
From this expression, it can be seen that a pair of Majorana fields with $m_L$ = $m_R$ and $\psi_L = i\sigma^2\psi_R^*$ is equivalent to a Dirac Field. This 4-component form will be of use in understanding the different issues concerning mass in the following sections.

Further insight into the relation between Dirac and Majorana fermions can be gained by quantizing the theory defined by the Dirac equation (\ref{Eqn:DirEqn}). To this end one can write the field operator for the Dirac fermion 
\begin{equation}
\label{field1}
\hat{\Psi}(x) =\sum_{E>0}a_Ee^{-iEt}\Phi_E(\bx)+\sum_{E<0}b^\dagger_{-E}e^{-iEt}\Phi_{E}(\bx),
\end{equation}
where $\Phi_E(\bx)$ is an eigenstate of the stationary Dirac equation at energy $E$ while $a^\dagger_E$, $b^\dagger_E$ are creation operators\footnote{
For readers unfamiliar with the formalism of second quantization we remark that  $a_E$, $b_E$ may be regarded as ordinary $c$-number coefficients.  Eqs.\ (\ref{field1}) and (\ref{field4}) then give expressions for the time-dependent wavefunction solutions of the Dirac equation representing the Dirac and the Majorana fermion, respectively. 
}
 for the particle and antiparticle with energy $E$, respectively. They satisfy the canonical fermionic anticommutation rules indicated in Eq.\ (\ref{can0}). By reversing the sign of the dummy summation variable in the second term and using the charge conjugation property Eq.\ (\ref{cc}) we may recast this as a sum over positive energy modes, 
\begin{equation}
\label{field2}
\hat{\Psi}(x) =\sum_{E>0}\left[a_Ee^{-iEt}\Phi_E(\bx)+b^\dagger_{E}e^{iEt}C\Phi^*_{E}(\bx)\right].
\end{equation}
The Dirac field operators can now be easily shown to satisfy the following characteristic equal-time anticommutation relations
\begin{eqnarray}
\label{field3}
\{\hat{\Psi}_a(x),\hat{\Psi}_b^\dagger(x')\}&=&\delta_{ab}\delta(\bx-\bx'), \nonumber\\  
\{\hat{\Psi}_a(x),\hat{\Psi}_b(x')\}&=&0,
\end{eqnarray}
where $a,b=1\cdots 4$ label their spinor components. Also, since obviously $\hat{\Psi}^\dagger(x)\neq\hat{\Psi}(x)$, we conclude that in this construction particles are distinct from antiparticles. The corresponding second quantized Dirac Hamiltonian ${\cal H}_D=\int d^3x\hat{\Psi}^\dagger(x)({\bm \alpha}\cdot{\bm p}+m_D\beta)\hat{\Psi}(x)$ is of the form
\begin{equation}
\label{field_hd1}
{\cal H}_D =\sum_{E>0} E (a_E^\dagger a_E+b^\dagger_E b_E).
\end{equation}

To construct the field operator for the Majorana fermion we demand that $a_E=b_E$ in Eq.\ (\ref{field2}), obtaining
\begin{equation}
\label{field4}
\hat{\Psi}(x) =\sum_{E>0}\left[a_Ee^{-iEt}\Phi_E(\bx)+a^\dagger_{E}e^{iEt}C\Phi^*_{E}(\bx)\right].
\end{equation}
The meaning of Eq.\ (\ref{field4}) is easiest to grasp when one employs the Majorana representation of the Dirac matrices. As we already noted the charge conjugation matrix is simply $C=\mathds{1}$ in this case. For the anticommutation algebra we then obtain 
\begin{eqnarray}
\label{field5a}
&&\hat{\Psi}^\dagger_a(x)=\hat{\Psi}_a(x), \\ &&\{\hat{\Psi}_a(x),\hat{\Psi}_b(x')\}=\delta_{ab}\delta(\bx-\bx'). 
\label{field5b}
\end{eqnarray}
The first equation informs us that, at the level of field operators, a Majorana particle is indistinguishable from its antiparticle. Alternately, the nonzero right hand side of the second equation can be taken as a defining property of the Majorana particle. In another representation, when $C$ differs from unity, Eq.\ (\ref{field5a}) is modified to $C_{ab}\hat{\Psi}_b^\dagger(x)=\hat{\Psi}_a(x)$ and $\delta_{ab}\to C_{ab}$ in the anticommutator, but the physical content remains the same. In the Majorana case the Hamiltonian becomes
\begin{equation}
\label{field_hd2}
{\cal H}_M =\sum_{E>0} E a_E^\dagger a_E,
\end{equation}
and the system can be seen to contain half as many independent degrees of freedom as ${\cal H}_D$. 

The Dirac equation of the particle physics thus allows for two
fundamentally different types of solutions describing a massive  spin-${1\over 2}$ particle. The original Dirac fermion can carry electrical charge and is distinct from its antiparticle; the two are related by the charge conjugation symmetry $\cal C$. Mathematically, this is obtained from an ``unconstrained'' solution of the Dirac equation. The Majorana solution of the same equation is obtained by imposing a reality constraint on the time-dependent wavefunction and describes a truly neutral spin-${1\over 2}$ particle indistinguishable from its antiparticle. In the field theory formulation both Dirac and Majorana fermions can be constructed from the same unconstrained solution of the stationary Dirac equation but the reality  constraint must then be imposed on the Majorana field at the operator level. The two descriptions are physically equivalent.

\subsection{Majorana Fermions as Emergent Particles in Solids with
  Superconducting Order}

As already noted, the Majorana equation also arises naturally in the
description of electrons in solids with superconducting order. This
section outlines how this comes about and points out similarities and
differences with the original Majorana theory.

To set the stage we first briefly review the Bogoliubov-de Gennes (BdG)
formalism \cite{deGennes1966} that is used to describe solids with superconducting
order. This is, in essence, the venerable BCS theory of
superconductivity \cite{BCS1957}, adapted to describe spatially non-uniform
situations. We use this formalism to elucidate how Majorana fermions
can arise in superconductors on general grounds. In later subsections
we then give specific examples of this general principle and connect
these examples to the ongoing experimental studies.   

Superconductivity arises when electrons in a metal
experience attractive interaction. In ordinary superconductors this
attraction is known to originate from the electron-phonon coupling but
other mechanisms have been proposed to operate in high-$T_c$ cuprate
and other unconventional superconductors. For our purposes the origin
of the attraction will be unimportant and we describe the superconductor
by the following minimal model,
\begin{equation}\label{hh1}
{\cal H}=\int d^dr\left[
h_0^{\sigma\sigma'}(\br)c^\dagger_{\sigma\br}c_{\sigma'\br}
- V n_{\uparrow\br}n_{\downarrow\br}\right].  
\end{equation}
Here $c^\dagger_{\sigma\br}$ creates an electron with spin $\sigma$ at
the spatial point $\br$ and
$n_{\sigma\br}=c^\dagger_{\sigma\br}c_{\sigma\br}$ denotes the
number operator. The first term in Eq.\ (\ref{hh1}) describes the kinetic
energy of the electrons and any single-electron potential while the second term represents the attractive
interaction with  $V>0$. In the simplest case of free electrons
$h_0^{\sigma\sigma'}(\br)=
(-\hbar^2\nabla^2/2m_{\rm eff}-\mu)\delta_{\sigma\sigma'}$ where $m_{\rm eff}$ represents
the electron band mass and $\mu$ the chemical potential.

To proceed we now perform the Bogoliubov mean-field decoupling of the
interaction term, writing
\begin{eqnarray}\label{deco}
-n_\uparrow n_\downarrow &=& 
c^\dagger_\uparrow c^\dagger_\downarrow c_\uparrow c_\downarrow \\
&\simeq& \langle c^\dagger_\uparrow c^\dagger_\downarrow \rangle
c_\uparrow c_\downarrow 
+c^\dagger_\uparrow c^\dagger_\downarrow \langle c_\uparrow
c_\downarrow \rangle 
-\langle c^\dagger_\uparrow c^\dagger_\downarrow \rangle \langle c_\uparrow
c_\downarrow \rangle, \nonumber
\end{eqnarray}
where the expectation values are taken with respect to the BdG mean-field
Hamiltonian specified below and we have suppressed the spatial index
for brevity. If we now
define the superconducting (SC) order parameter 
\begin{equation}\label{del1}
\Delta(\br)=V\langle c_{\uparrow\br}c_{\downarrow\br} \rangle,
\end{equation}
we can write down the BdG mean-field Hamiltonian
\begin{eqnarray}\label{hh2}
{\cal H}_{\rm BdG} &=& \int d^dr\biggl[
h_0^{\sigma\sigma'}(\br)c^\dagger_{\sigma\br}c_{\sigma'\br} \\
&+& \left(\Delta(\br)c^\dagger_{\uparrow\br} 
c^\dagger_{\downarrow\br}+{\rm h.c.}\right) \nonumber 
-{1\over V}|\Delta(\br)|^2 \biggr].  
\end{eqnarray}
As the final step we define a four-component Nambu spinor
\begin{equation}\label{hh2a}
\hat\Psi_\br=
\begin{pmatrix}
c_{\uparrow\br} \\ c_{\downarrow\br} \\ c^\dagger_{\downarrow\br} \\ -c^\dagger_{\uparrow\br}
\end{pmatrix}
\equiv
\begin{pmatrix}
\hat{\psi}_\br \\ i\sigma^y\hat{\psi}^*_\br
\end{pmatrix},
\end{equation}
where the hat symbol reminds us that $\hat\Psi_\br$ is an operator.
This allows the BdG Hamiltonian to be cast into a compact form
\begin{equation}\label{hh3}
{\cal H}_{\rm BdG} = \int d^dr\biggl[
\hat\Psi^\dagger_\br H_{\rm BdG}(\br) \hat\Psi_\br
-{1\over V}|\Delta(\br)|^2 \biggr],
\end{equation}
with 
\begin{equation}\label{hbdg1}
 H_{\rm BdG}(\br) =
\begin{pmatrix}
h_0(\br) & \Delta(\br) \\
\Delta^*(\br) & -\sigma^y h^*_0(\br) \sigma^y
\end{pmatrix}.
\end{equation}
In the last equation $h_0$ and $\Delta$ should be viewed as $2\times
2$ matrices in spin space while $\bsig=(\sigma^x,\sigma^y,\sigma^z)$ denotes the
corresponding vector of Pauli matrices. It is useful at this point to
introduce another set of Pauli matrices
${\bm\tau}=(\tau^x,\tau^y,\tau^z)$ acting in the Nambu space, i.e.\
the $2\times 2$ matrix structure explicitly displayed in Eq.\
(\ref{hbdg1}). We also note that the size of the Hamiltonian matrix had to be doubled to accommodate the pairing term. As a result, only half of its independent solutions are physical.

The problem specified by the Hamiltonian (\ref{hh3}) and the
self-consistency condition (\ref{del1}) can now be solved by seeking
a set of eigenfunctions
$\Phi_n(\br)=[u_{n\uparrow}(\br),u_{n\downarrow}(\br),v_{n\uparrow}(\br),v_{n\downarrow}(\br)]^T$
and eigenvalues $E_n$ satisfying the stationary BdG equation
\begin{equation}\label{bdg1}
 H_{\rm BdG}(\br)\Phi_n(\br) = E_n\Phi_n(\br).
\end{equation}
In the basis spanned by these eigenfunctions the Hamiltonian (\ref{hh3})
 is brought to a diagonal form,
\begin{equation}\label{hh4}
{\cal H}_{\rm BdG} = {\sum_n}' E_n a^\dagger_n a_n+E_g
\end{equation}
where $E_g$ is a constant representing the ground-state energy and the prime indicates that the summation is restricted to positive energies $E_n$ to avoid double-counting of modes resulting from the doubled matrix size.
\begin{equation}\label{psi1}
a_n=\int d^dr\Phi_n^\dagger(\br)\hat\Psi_\br
\end{equation}
is the eigenmode operator that annihilates the Bogoliubov quasiparticle
with energy $E_n$ and satisfies the canonical fermionic anticommutation algebra (\ref{can0}). 


The connection between the BdG theory and the Majorana construction is most directly apparent in the structure of the Nambu spinor $\hat{\Psi}_\br$ defined in Eq.\ (\ref{hh2a}). This is nothing but an operator version of Eq.\ (\ref{Eqn:PsiLR}). The Nambu spinor satisfies the Majorana condition (\ref{ccm}): indeed it holds
\begin{equation}\label{ph0}
C\hat{\Psi}^*_\br=\hat{\Psi}_\br,
\end{equation}
where $\hat{\Psi}^*_\br=(\hat{\Psi}_\br^\dagger)^T$ and 
$C=\tau^y\sigma^y$
is the charge conjugation matrix. One could furthermore construct the field operator $\hat{\Psi}(t,\br)$ of the BdG theory which would have the same structure as the Majorana field operator in Eq.\ (\ref{field4}). This property follows directly from the fact that the lower two components of the Nambu spinor $\hat{\Psi}_\br$ are related to the upper two, as indicated in Eq.\ (\ref{ph0}) above. The form of the Nambu spinor is in turn dictated by the structure of the second quantized BdG Hamiltonian (\ref{hh2}). So, unlike with the Dirac equation, where the choice between Dirac and Majorana solutions is ours to make, the theory of superconductors requires description in terms of Majorana particles.  

Finally, we note that the BdG Hamiltonian in the eigenmode representation (\ref{hh4}) has the same form as the Majorana Hamiltonian ${\cal H}_M$ defined in Eq.\ (\ref{field_hd2}).


Another similarity with particle physics follows from the anomalous
terms $c^\dagger_{\uparrow\br} c^\dagger_{\downarrow\br}$ appearing in
the BdG Hamiltonian (\ref{hh2}). These indicate events in which the
number of electrons changes by $\Delta L=2$, analogous to the lepton
number non-conservation discussed in Sec.\ \ref{Sec:MajFermNucPart}C
below. In the superconductor, the total number of electrons is of
course strictly conserved, as one can see by inspecting the full
interacting Hamiltonian (\ref{hh1}) which commutes with the number
operator. In the BdG description the number non-conservation reflects
the fact that a pair of electrons can disappear (or emerge from) the
superconducting condensate which is treated at the mean-field
level. Such processes are akin to Majorana pair annihilation in
particle physics and have experimentally observable consequences \cite{Beenakker2014}.  

One can thus say that in the BdG theory defined by the
Hamiltonian ${H}_{\rm BdG}(\br)$ the excitations of a superconductor
possess all the key attributes of Majorana fermions: they are electrically
neutral fermions with no distinction between particles and
antiparticles. The SC gap $\Delta$ plays the role of the Majorana mass. This point of view has been
generally appreciated for a long time but was carefully analyzed and
emphasized only fairly recently \cite{Chamon2010}.

It may be concluded from the analysis presented above that Majorana
fermions naturally appear in the theoretical description of a
generic superconductor. Since the  BCS theory and the related BdG
formalism on which this description rests are in an excellent agreement
with the large body of experimental data on superconductors it could be
inferred that the existence of Majorana fermions is already well established in this
context. 

The recent interest in condensed matter physics has been centered around Majorana zero modes (MZMs), already briefly mentioned in connection with the Majorana condition (\ref{ccm}). MZMs constitute a special case of Majorana fermions that occur at exact zero energy and are typically localized in space in the vicinity of defects, such as vortices or domain walls. Their key property is that the stationary state associated with the MZM by itself satisfies the Majorana condition. Such zero modes occur in so called topological superconductors and in fact do not obey the ordinary fermionic exchange statistics but behave as ``non-Abelian anyons'' \cite{Moore1991, Nayak2008}.
It is this extremely interesting property that has motivated intense theoretical and experimental studies over the
past decade and made searches for MZMs among the most active subfields of
condensed matter physics. We devote Sec.\ IV to the detailed discussion of Majorana zero modes, topological superconductors and the explanation of the of non-Abelian exchange statistics. It is important to emphasize that these phenomena associated with MZMs in solids only occur in one- and two-dimensional systems and have no direct analog in high energy physics.

\section{Majorana Fermions in Nuclear and Particle Physics}
\label{Sec:MajFermNucPart}

The neutrino is the {\em usual suspect} when one discusses fundamental particles that might be Majorana in nature. The neutrino only interacts weakly and therefore it is very difficult to observe its behavior. As a result, there are key aspects of the neutrino, such as its mass, that are still unknown. Furthermore, the weak interaction violates parity and therefore right-handed neutrinos (and left handed anti-neutrinos) have no interaction. Therefore, it is unknown if those states are unobservable or simply don't exist. The quantum mechanics of neutrinos, be they Dirac or Majorana particles, has been described in detail in many places. (See for example Refs.~\cite{Boehm1987,Kayser1989,Mohapatra1991} and \cite{Zralek1997}.) Here we provide an overview.

\subsection{The Seesaw Mechanism}
\label{sec:seesaw}
In field theory, the wave equations given by Eqs.~\ref{Eqn:DirEqn} and~\ref{Eqn:MajEqn} can be derived from a Lagrangian density ($\mathcal{L}$) using a variational principle (the Euler-Lagrange equation). The appropriate $\mathcal{L}$ includes mass terms ($\mathcal{L}_{m}$) whose form would depend on whether the particles were described as Majorana or Dirac. In the case of neutrinos, for example, it is not known which type they are so in principle both possibilities should be considered. Although there are 3 flavors of neutrino, it is instructive to look at $\mathcal{L}_{m}$ for a lone flavor. Hence the mass terms of $\mathcal{L}_{m}$ are written (where h.c. is shorthand for hermitian conjugate):
\begin{eqnarray}
\label{Eqn:Langr}
\mathcal{L}_{m} &=& m_D[\bar{\nu}_R\nu_L + (\bar{\nu}_L)^c\nu_R^c]  \\ \nonumber
& +& \frac{1}{2}m_L[(\bar{\nu}_L)^c\nu_L + \bar{\nu}_L\nu_L^c]+ \frac{1}{2}m_R[(\bar{\nu}_R)^c\nu_R + \bar{\nu}_R\nu_R^c] \\ \nonumber
 &=& \frac{1}{2}(\begin{array}{c}(\bar{\nu}_L)^c ~~ \bar{\nu}_R\end{array})\left(\begin{array}{cc} m_L & m_D \\ m_D & m_R \end{array}\right)\left( \begin{array}{c}\nu_L \\ ({\nu}_R)^c \end{array} \right) + {\mbox h.c.}
\end{eqnarray}
where we have introduced the notation $\nu_{R,L}$ for the respective neutrino annihilation operators replacing the generic $\Psi_{R,L}$ notation of Eq.~(\ref{Eqn:PsiLR}). With the possibility that all three mass terms exist, $\mathcal{L}_{m}$ must be diagonalized resulting in two mass eigenvalues and their corresponding eigenstates. If $m_L=m_R=0$, the result is a lone neutrino described by a 4-component Dirac spinor with two equal masses, one for particle and one for anti-particle. In the seesaw model~\cite{Yanagida1979,GellMann1979,Minkowski1977,Mohapatra1980}, the assumption is made that $m_R$ is very large compared to $m_D$ and $m_L$ is zero. This was motivated by the non-observation of $\nu_R$, which can be explained if its mass is very heavy. It also seems natural that $m_D$ should have a value near that of the charged fermions (e.g. the electron). Under these assumptions, diagonalizing the matrix one finds that the two eigenvalues are $m_R$ and $m_{\nu} \sim m_D^2/m_R$. Its critical to notice here that $m_{\nu}$ is much smaller than the typical charged Dirac lepton, which agrees with the important empirical fact that neutrino masses are much less than those of their charged partners. Equally important, in this case, is that the diagonalization results in 2 Majorana neutrinos, each described by a 2-component spinor, where one is light and one heavy. This seesaw mechanism not only provides a hint as to why neutrino masses are so small but also ties that hint to the character of neutrinos being Majorana.

\subsection{Lepton Number Conservation}
Empirically, no process has been observed that changes the total number of leptons. This fact is usually stated as lepton number ($L$) being conserved. Consider the interactions given in Eq.~(\ref{eqn:NeutronDecay}). In the expression for neutron decay, the neutron and proton are baryons with $L=0$. The beta particle has $L=1$ and the anti-neutrino has $L=-1$. Before and after the decay, the total number of leptons is 0 and therefore $L$ is conserved. This is also similar for the inverse beta decay reaction, where $L=1$ before and after the interaction.
\begin{eqnarray}
\label{eqn:NeutronDecay}
n 	&	\rightarrow 		&	p + \beta^- + \overline{\nu}_e~ \\ \nonumber
\nu_e + n 	& \rightarrow 	&	p + \beta^- 
\end{eqnarray}
Terms such as $(\bar{\nu}_L)^c\nu_L$ in Eq.~(\ref{Eqn:Langr}) result in interactions, such as double beta decay (\ref{sec:doublebeta}) that change $L$ by 2 units. That is, $L$ would not be conserved. Therefore, due to such an interaction, the $\overline{\nu}_e$ produced in beta decay could initiate the inverse beta decay reaction with the result that $\Delta L = 2$. In fact, this pair of processes was actually one of the early proposed tests of Majorana neutrinos~\cite{Pontecorvo1957,Pontecorvo1958}. 

\begin{figure}[!htbp]
\includegraphics[width=6cm]{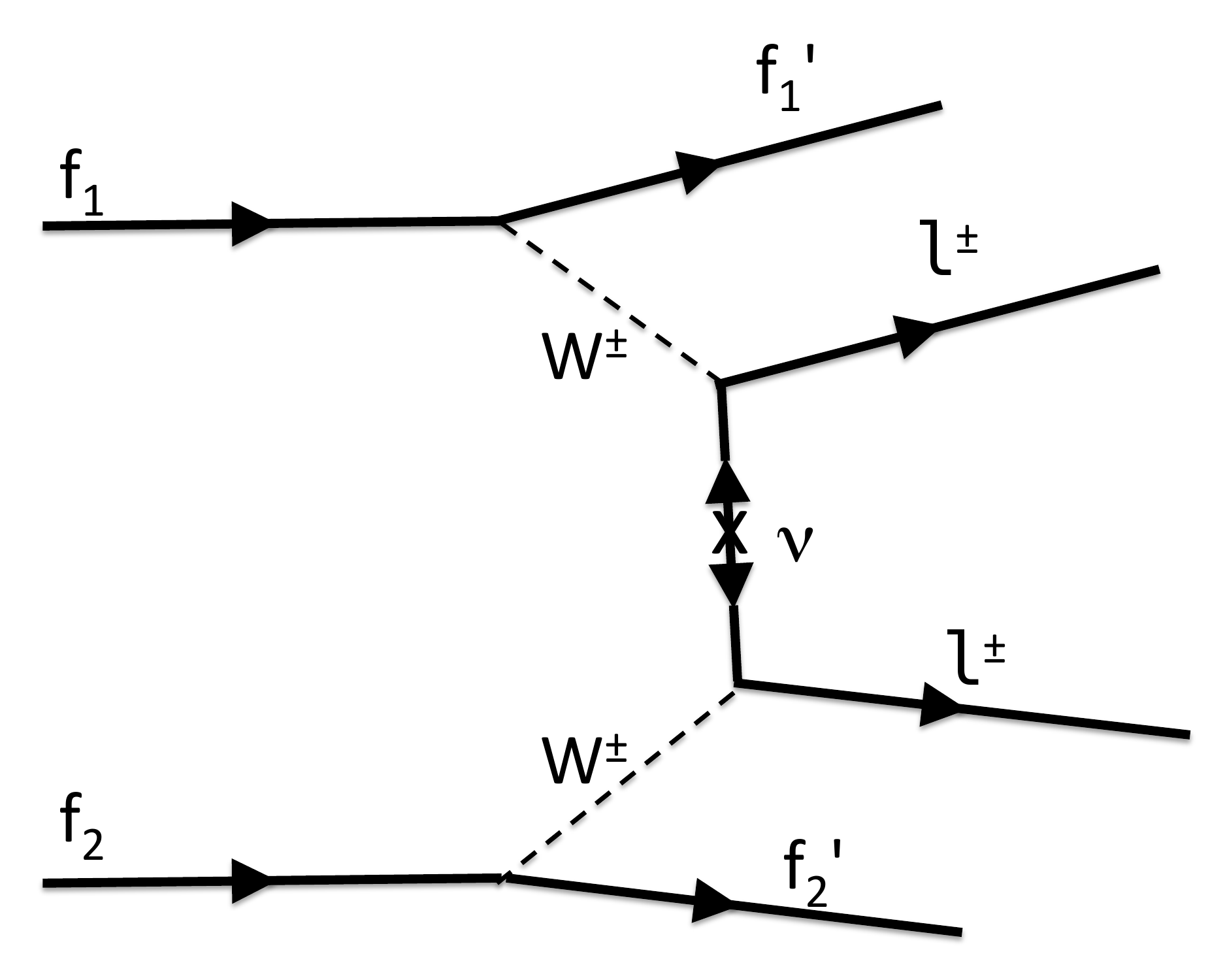}
\caption{Generic diagram showing a $\Delta L=2$ process involving the exchange of a Majorana neutrino. $f$ is a generic fermion, $W^{\pm}$ is the weak interaction intermediate vector boson, and $l^{\pm}$ is an outgoing lepton. The two $W$ and $l$ have the same charge for this $\Delta L=2$ process.}
\label{fig:GenericFeynmann}
\end{figure}

One important feature of the seesaw mechanism is the prediction of not only light, but also heavy Majorana neutrinos. There are numerous searches for both such particles through the $\Delta L=2$ processes they would mediate. Figure~\ref{fig:GenericFeynmann} displays a generic $\Delta L=2$ process involving an exchanged Majorana neutrino.

\subsection{Leptogenesis}
\label{sec:leptogenesis}
The seesaw model described above, not only gives insight into the smallness of $\nu$ mass, it also provides a mechanism to explain the matter-antimatter asymmetry of the Universe. This mechanism is called leptogenesis~\cite{Fukugita1986} and recent reviews are available~\cite{Buchmuller2005,Davidson2008,DiBari2012}.

The general requirements for a dynamical process to produce this asymmetry were identified in 1967 by A. Sakharov ~\cite{Sakharov1967}. First, the conservation of baryon number ($B$), the number of protons and neutrons, must be violated in some process. Second, charge-conjugation and space-inversion ($CP$) conservation must also be violated. If $CP$ is conserved, then the rate of a process will be identical to the related process where all particle charges are reversed (i.e. $Q \rightarrow -Q$) and spatial coordinates are inverted (i.e. $\bm{r} \rightarrow -\bm{r}$).  Finally, the processes that violate these conservation laws must take place out of equilibrium.

Majorana neutrinos violate $L$. This net $L$ can be converted to a net $B$ through standard model processes. The mass matrices represented in Eq.~\ref{Eqn:Langr} have phases that may lead to $C$ and $CP$ violation. Finally, the heavy Majorana neutrinos have no gauge interactions and therefore can fall out of thermal equilibrium with the other particles in the primordial soup. Thus Majorana neutrinos and leptogenesis provide all of the Sakharov requirements. As such Leptogenesis is presently an active area of research into the origin of the matter, anti-matter asymmetry. If neutrinos are shown to be Majorana, it will be yet another clue as to our origins.

\subsection{Double Beta Decay}
\label{sec:doublebeta}
The search for double beta decay (\BB) is primarily motivated by its ability to demonstrate that neutrinos are Majorana, if that happens to be the case. It is a second order, weak process closely related to beta decay. Most nuclei that have both an even number of protons and neutrons are stable against beta decay. That is, the process
\begin{equation}
^N_ZA \rightarrow ^{N-1}_{Z+1}A + \beta^- + \overline{\nu}_e~,
\end{equation}
is energetically forbidden in most nuclei of mass number $A$ that have an even atomic number $Z$ and even neutron number $N$. In even-even nuclei for which beta decay is allowed, the rate is greatly inhibited. For a large number of even-even nuclei, however, the second order process with $Z$ changing by 2 units, while emitting 2 electrons and 2 anti-neutrinos is allowed:
\begin{equation}
^N_ZA \rightarrow ^{N-2}_{Z+2}A + 2\beta^- + 2\overline{\nu}_e~.
\end{equation}
This process is called two-neutrino double beta decay (\BBt). As the 2 $\beta^-$s are leptons and the 2 $\overline{\nu}_e$s are anti-leptons, the total lepton number before and after the decay is unchanged and $L$ is conserved. \BBt\ is expected within the standard model and has been observed in about 10 isotopes~\cite{Barabash2010,Barabash2013}. 

\begin{figure}[!htbp]
\includegraphics[width=6cm]{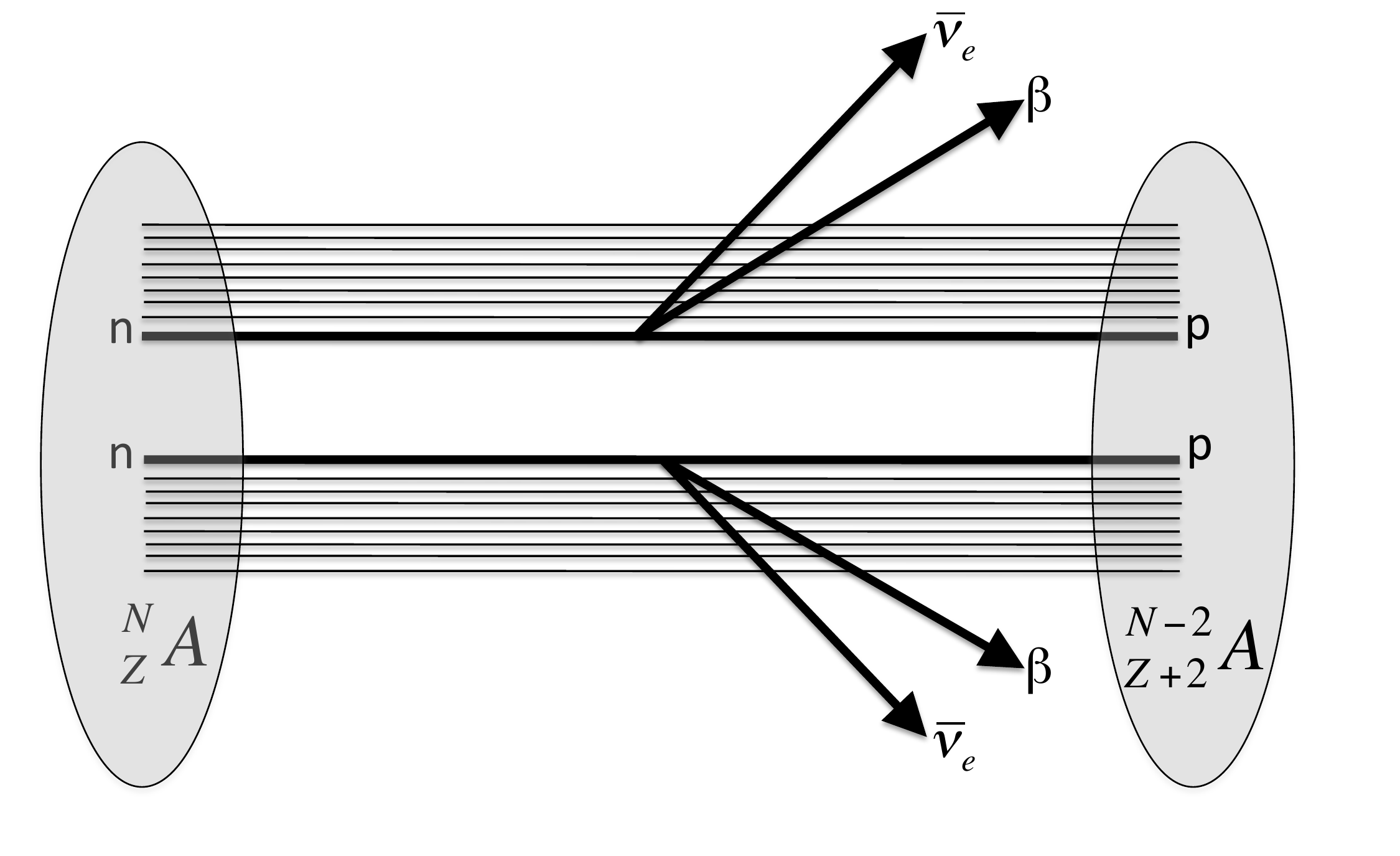}
\includegraphics[width=6cm]{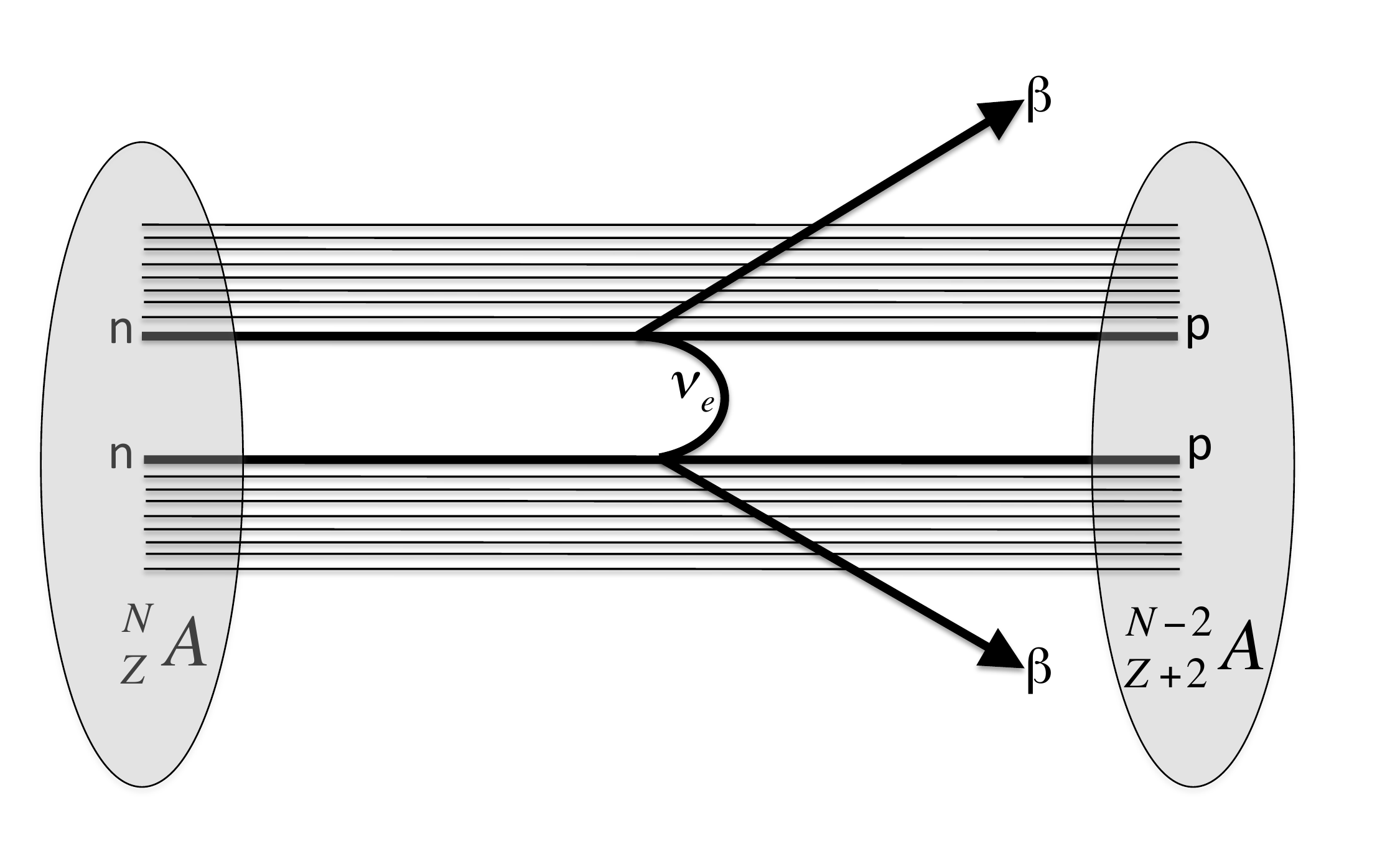}
\caption{Diagrams showing the \BBt\ (top) and \BBz\ (bottom) processes. Within the group of nucleons inside a nucleus, two neutrons simultaneously emit $\beta$ particles while producing protons.}
\label{fig:BBFeynmann}
\end{figure}

An alternative process that emits no neutrinos is written
\begin{equation}
^N_ZA \rightarrow ^{N-2}_{Z+2}A + 2\beta^-~.
\label{eqn:BB0nu}
\end{equation}
This zero-neutrino, or neutrinoless double beta decay (\BBz) process violates lepton number by 2 units. Diagrams representing \BBt\ and \BBz\ are shown in Fig.~\ref{fig:BBFeynmann}. Note the common features of the \BBz\ panel in this figure to that of Fig.~\ref{fig:GenericFeynmann}.

\begin{figure}[!htbp]
\includegraphics[width=6cm]{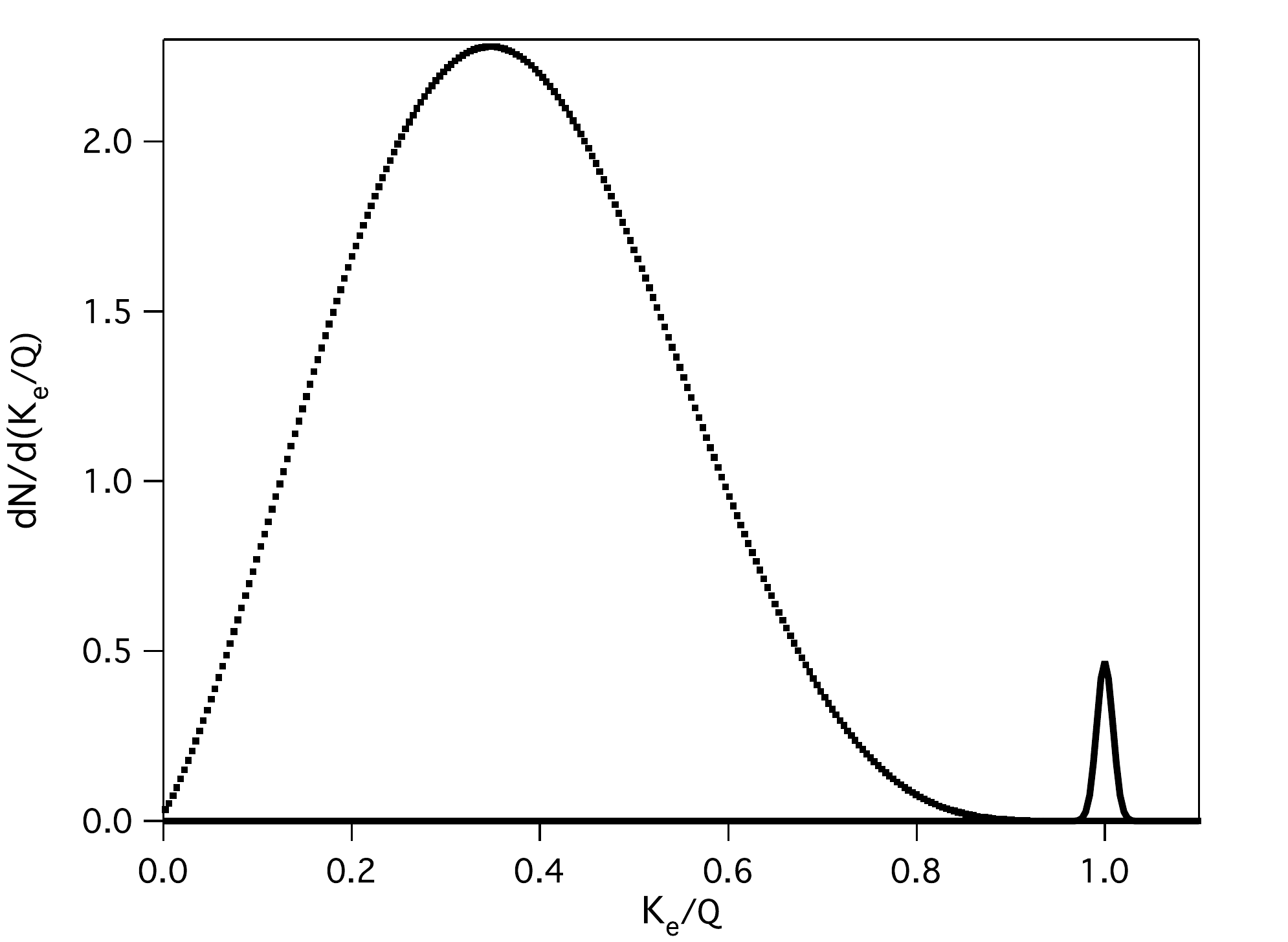}
\caption{The spectrum of the sum of the energies of the two electrons from \BBt\ (dotted) and \BBz\ (solid). The resolution used for \BBz\ is arbitrary and the relative strength of \BBz\ to that of \BBt\ is exaggerated for clarity. In reality, if \BBz\ exists, the peak would be very weak.}
\label{fig:BBSpectra}
\end{figure}

In experiments designed to directly detect \BB, the two decay modes are distinguished by the energy carried off by the exiting $\beta$ particles. (Section~\ref{Sec:BBexpt} discusses the various experimental issues.) The spectra are shown in Fig.~\ref{fig:BBSpectra}. Since the neutrinos interact too weakly for their energy to be feasibly observed, the \BBt\ spectrum of the sum of the electron energies is a continuum up to the total available energy for the decay. In contrast, the sum of the energies of the two electrons in \BBz\ is a mono-energetic peak at that endpoint.

It may seem ironic that one can learn about neutrinos from a process that produces no neutrinos. The key to understanding this apparent non sequitor lies in Fig.~\ref{fig:BBFeynmann}. In the standard model, when a neutron decays it emits an $\overline{\nu}_e$, whereas a neutron can only absorb a $\nu_e$. If the neutrino and anti-neutrino are distinct particles, the exchange depicted in the the lower panel of that figure cannot occur. If the exchange does take place, there is no such distinction and neutrinos must be Majorana particles.

Figure~\ref{fig:BBFeynmann} depicts \BBz\ proceeding through the exchange of a light mass neutrino. It is known that neutrinos have a small mass~\cite{Fogli2012}, although the magnitude of that mass is not yet known. Since we know light neutrinos exist, this light-neutrino exchange is the most commonly considered mechanism for \BBz\ in the literature. Numerous other possibilities have been proposed over the years. (See Ref.~\cite{Gehman2007} and references therein for a list.) It should, however, be mentioned that if some other mechanism does mediate the decay, an observation of \BBz\ still implies that neutrinos are massive Majorana particles as shown by the Schechter-Valle theorem~\cite{Schechter1982}\footnote{The Schechter-Valle theorem however, does not ensure that we can deduce a unambiguous value for the neutrino mass given a measurement of \BBz. The theorem shows that massive Majorana neutrinos exist, but the light-neutrino-mass contribution to the decay rate could be sub-dominate~\cite{Duerr2011}.}.

The decay rate for \BBz\ can be written:
\begin{equation}
\label{eq:BBrate}
[\mbox{\Tz}]^{-1} = \mbox{\Gz} |\mbox{\Mz}|^2 \mbox{\mee}^2
\end{equation}
where \Tz\ is the half-life of the decay, \Gz\ is the kinematic phase space factor, \Mz\ is the matrix element corresponding to the \BBz\ transition, and \mee\ is the effective Majorana neutrino mass. \Gz\ contains the kinematic information about the final state particles, and is  calculable to the precision of the input parameters. \Mz\ is difficult to calculate with an accuracy estimated to be approximately a factor of 2. One immediately notices from Eq.~(\ref{eq:BBrate}) that the decay rate is directly related to the Majorana neutrino mass. As the neutrino mass trends toward zero, the decay rate will also. In general, as the neutrino mass vanishes, it becomes impossible to discern the Dirac or Majorana nature of the neutrino.

The weak eigenstates of neutrinos, the quantum states produced during a weak interaction such as $\beta$ decay, are not identical to the mass eigenstates. As a consequence, neutrinos will {\em oscillate} between the weak eigenstates as they propagate through space. This empirical fact has permitted the use of interferometry to study many neutrino characteristics including the difference between the mass eigenvalues and the mixing matrix elements that describe the oscillations. (A review of neutrino oscillations can be found in Ref.~\cite{Balantekin2013}.) The value of \mee\ depends on these mixing angles and mass eigenvalues. If the neutrino is Majorana, then one can derive an expression for \mee\ that includes data from neutrino oscillation experiments. It is written
\begin{eqnarray}
\label{eqn:mee}
\mbox{\mee} & = & |\sum_{i=1}^3U_{ei}^2m_i|		 \\
 &=& |U_{e1}^2m_1 + U_{e2}^2m_2 + U_{e2}^2m_2|	\nonumber \\
 &=& |m_1c_{12}^2 c_{13}^2 + m_2s_{12}^2 c_{13}^2e^{i\alpha_{21}} + m_3s_{13}^2 e^{i\alpha_{31}}|. \nonumber 
\end{eqnarray}
where $m_i$ are the mass eigenvalues and the $U_{ei}$ are the mixing matrix elements. The final line of Eq.~(\ref{eqn:mee}) expresses the mixing matrix element in terms of mixing angles with the notation $c_{12} \equiv cos\theta_{12}$. There are two phases that appear ($\alpha_{21},\alpha_{31}$) that arise because of the Majorana nature of the neutrino. 

Neutrino oscillation experiments are insensitive to the Majorana-Dirac nature of the neutrino. Importantly, these experiments also cannot determine the absolute mass scale of the neutrino.  Such experiments only determine mass differences and hence a relationship between the three $m_i$. (To be technically correct, these experiments determine the differences in the masses squared.) As a result, the oscillation experiments have some, but not perfect, predictive power for the value of \mee, and therefore \Tz, if neutrinos are Majorana particles. There is a region of \mee\ between about 15 and 50 meV, where an optimist would expect to see \BBz. This region is referred to as the inverted-hierarchy or atmospheric mass region. It originates if the dominate contributions to \mee\ arise from the heavier mass values associated with oscillation channels first observed in atmospheric neutrinos. A more pessimistic view is that \mee\ would be a few meV or less, the so-called normal-hierarchy or solar mass region. This mass region arises from the lighter mass values associated with oscillation results first observed in solar neutrinos.

\begin{figure}[!htbp]
\includegraphics[width=8cm]{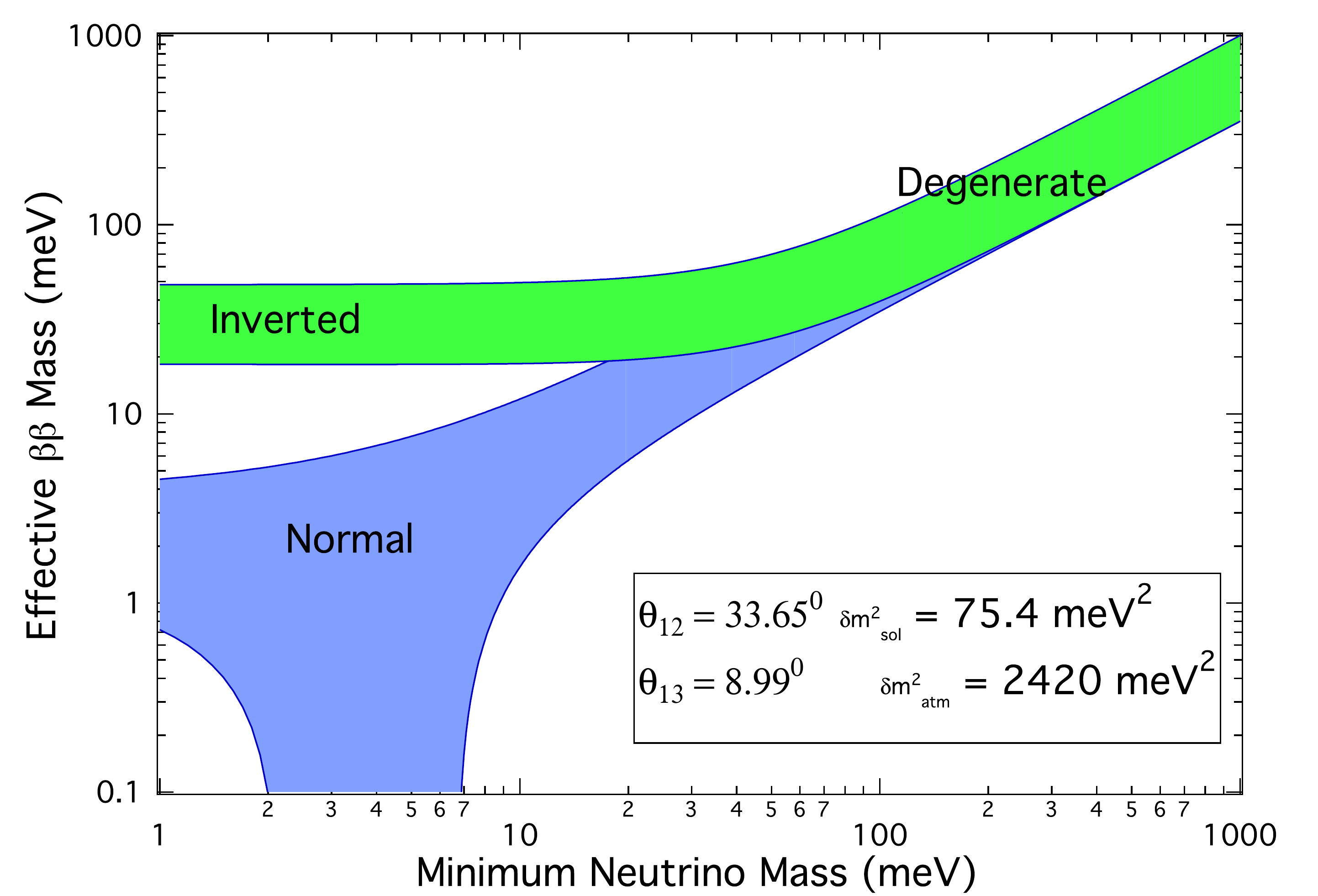}
\caption{\mee\ as a function of the smallest of the three mass eigenvalues. The plot is done for the best fit oscillation parameters~\cite{Fogli2012}.}
\label{fig:MinMax}
\end{figure}

If one makes a plot of \mee\ versus the lightest neutrino mass (Fig.~\ref{fig:MinMax}) using what is known about the angles and mass differences, the result will be a band, as opposed to a line, due to the unknown phases. One will also see that two bands appear in such a figure, one for the normal and one for the inverted hierarchy that blend as one approaches a regime where all three mass eigenvalues are roughly equal, the degenerate region.

For the inverted hierarchy region, \Tz\ is expected to be about $10^{27}$ y. From Eq.~(\ref{eq:BBrate}), it is clear that if \mee\ is a factor of 10 less, that is within the normal hierarchy region, then \Tz\ would be near $10^{29}$ y. The next generation of experiments (Sec.~\ref{Sec:BBexpt}) are aiming to explore the inverted hierarchy region, with longer term R\&D aimed at exploring the normal hierarchy region if nothing is observed at the shorter half-life.

\subsection{Supersymmetry}
The Standard Model (SM) has been very successful describing observed phenomena up to an energy scale of about a TeV. Even so, there are a number of deficiencies. These include the existence of dark matter, neutrino masses, the matter-antimatter asymmetry, and why the interaction strength of the known forces varies so greatly (the hierarchy problem). 

Supersymmetry~\cite{Aitchison2009} (SUSY) is a proposed extension to the SM that would address these shortcomings. SUSY relates bosons to fermions through a new symmetry that joins a fermion (boson) into a supermultiplet with a boson (fermion). When all the known particles are placed into supermultiplets, one finds that no known particles are available to fill the role of the superpartners. That is, half the particles required by SUSY have yet to be observed.

Particles within such a supermultiplet share properties, although they have differing spins. The supermultiplet that contains the spin 1 photon will contain a new spin $\frac{1}{2}$ particle, referred to as the photino. Since the photon is its own antiparticle, so would be the photino. Since the photino is spin $\frac{1}{2}$, if it exists, it would a Majorana particle. In practice, there are 4 such new particles (the superpartners of the photon, Z and two Higgs bosons) that would mix with the possible combinations being called neutralinos, which would be Majorana particles. If one of these neutralino states is the lightest supersymmetric particle (LSP), it would be long-lived and might be the dark matter. Although other SUSY LSP candidates that are not Majorana in nature might also be dark matter candidates, SUSY provides for some Majorana dark matter candidates.

In general, SUSY predicts the existence of Majorana particles. The search for such particles is described in Sections~\ref{sec:AccerSearch} and~\ref{sec:DMSearch}.

\subsection{Prospects for Observation}
\label{Sec:ProspectsObserv}

Processes that violate total lepton number by two units ($\Delta  L = 2$) are indicators of Majorana neutrinos. In this section, we consider such processes and their prospects for discovering a Majorana particle. We treat \BBz\ first and in greater detail, as it is the most feasible technique to achieve this goal. Furthermore, the observation of \BB\ is an unambiguous signature for Majorana neutrinos~\cite{Schechter1982}. Most of the other Majorana particle searches discussed below would be indicative but not conclusive.  

Equation~\ref{eq:BBrate} makes it clear that the \BBz\ decay rate is directly related to the Majorana neutrino mass. Table~\ref{tab:PastExperiments} summarizes the most recent \BBz\ experimental results, which indicate \Tz\ is greater than $10^{25}$ y; more than $10^{15}$ times longer than the age of the Universe. This long half-life limit constrains the effective Majorana neutrino mass to be very small. Measuring, or placing limits on, such a slow process is possible because Avogadro's number is so large. Experiments to date have used about 10 moles of isotope, and future proposals will be much larger yet. It is the advantage from monitoring such a large number of atoms that makes \BBz\ the most sensitive technique to search for light Majorana neutrinos. Experiments that use neutrino sources and targets suffer from low event rates due to modest neutrino fluxes and small weak-interaction cross sections, when searching for $\Delta L = 2$ processes. Some accelerator searches for heavy Majorana neutrinos overcome these limitations when resonant interactions are considered. Hence, in certain limited mass regions, accelerator efforts can compete with \BBz.

\subsubsection{Double Beta Decay}
\label{Sec:BBexpt}

\begin{figure}[!htbp]
\includegraphics[width=8cm]{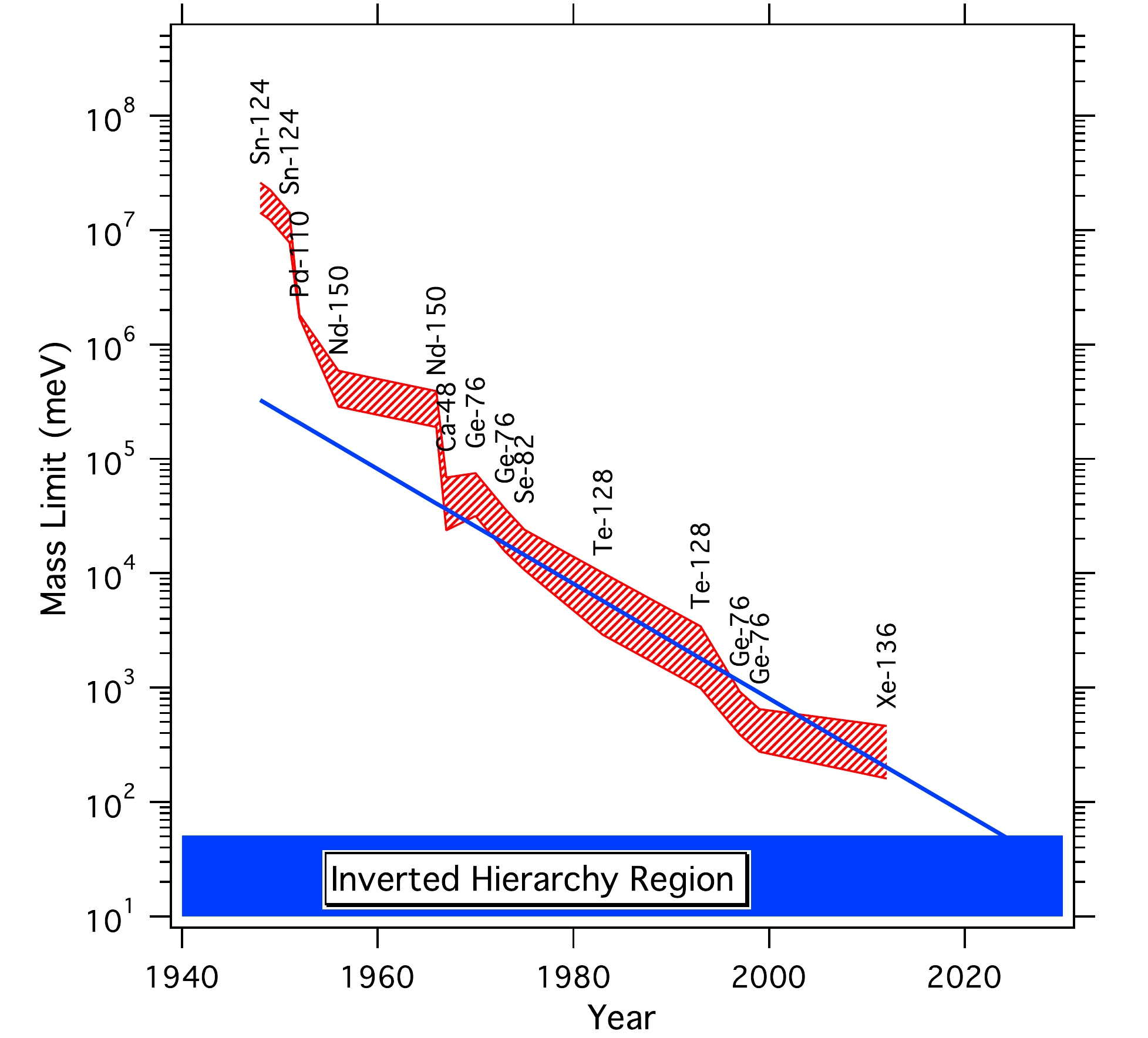}
\caption{A history of the effective Majorana neutrino mass limit from \BBz. The shaded band indicates the range of masses one would deduce from the reported \Tz\ as a result of the spread of matrix element calculations. The straight line is an extrapolation drawn by eye for a mass value improvement of a factor of 2 every 6 years. The lower shaded region is the target derived from neutrino oscillation experiments.}
\label{fig:BBHistory}
\end{figure}

Figure~\ref{fig:BBHistory} shows how the limit on \mee\ has evolved over the years. The limit improves by about a factor of 2 every 6 years. If this trend continues, the inverted-hierarchy goal for the Majorana mass sensitivity below 50 meV should be explored during the coming decade or so. Within the next few years, the presently operating experiments and those due to come online should extend the reach below 100 meV. The experimental and theoretical situation in \BB\ has been well summarized in numerous excellent reviews~\cite{ell02, ell04, bar04, Eji05, avi05, avi08, bar11, Rodejohann2011, Elliott2012, Vergados2012,Schwingenheuer2013}. In particular the current experimental program has been described in detail. Here we don't repeat that effort, but only summarize some highlights and direct the interested reader to the literature.

\begin{table*}[htdp]
\caption{A list of recent  \BBz\ experimental results and their 90\% confidence level limits on \Tz. The \mee\ limits are those quoted by the authors using the \Mz\ of their choice. The result on \nuc{76}{Ge}\cite{Agostini2013a} combines data from Ref.~\cite{kla01a,aal02a}.}
\begin{center}
\begin{tabular}{|c|c|c|c|c|}
\hline
\hline
Isotope                &  Technique                     & \Tz\                                 & \mee\ (eV)       		& Reference  \\
\hline
\nuc{48}{Ca}     & CaF$_2$ scint. crystals             &$>5.8\times10^{22}$ y               &   $<$3.5-22   			& \cite{Ume08}\\
\nuc{76}{Ge}   & \nuc{enr}{Ge} det.                      &    $>3.0 \times 10^{25}$ y          & $<$(0.2-0.4)           & \cite{Agostini2013a} \\
\nuc{82}{Se}    &Thin metal foils \& tracking               & $>3.2 \times 10^{23}$ y              & $<$(0.94-1.71)        &  \cite{Tre11}\\
\nuc{100}{Mo}   &Thin metal foils \& tracking               & $>1.1 \times 10^{24}$ y             & $< $(0.3-0.9)         &  \cite{Arnold2013}\\
\nuc{116}{Cd}   &\nuc{116}{Cd}WO$_4$ scint. crystals 		 & $>1.7 \times 10^{23}$ y              & $<$1.7       		& \cite{dane03}\\
\nuc{128}{Te}    & geochemical                           & $>7.7 \times 10^{24}$ y               & $<$(1.1-1.5)         & \cite{ber93} \\
\nuc{130}{Te}    & TeO$_2$ bolometers                     & $>2.8 \times 10^{24}$ y              & $<$(0.3-0.7)        &  \cite{arn08}  \\
\nuc{136}{Xe}   &  Liq. Xe scint.  						& $>1.9 \times 10^{25}$ y 			&  $<$(0.16-0.33) 		& \cite{Gando2013}    \\
\nuc{150}{Ne}   &   Thin metal foil within TPC    		&$>1.8 \times 10^{22}$ y               & N.A.                  & \cite{Bar10} \\
\hline
\hline
\end{tabular}
\end{center}
\label{tab:PastExperiments}
\end{table*}%

Recent results come from the EXO-200~\cite{Auger2012,Albert2014} and KamLAND-Zen~\cite{Gando2013} collaborations working with \nuc{136}{Xe} and the GERDA collaboration with with \nuc{76}{Ge}~\cite{Agostini2013a}.\footnote{These 3 results are in tension with an earlier claim~\cite{Klapdor2006} for the observation of \BBz.} Several additional experiments should have results by about 2016, including CUORE~\cite{Alessandria2011}, \MJ~\cite{Abgrall2013}, NEXT~\cite{Gom11}, SNO+~\cite{Hartnell2012}, and SuperNEMO~\cite{Barabash2012}. Half-life limits are presently beyond $10^{25}$ y with \mee\ limits below a couple hundred meV. Within a couple years, it is expected that the mass limit will reach 100 meV or less.

The sensitivity goal of the next generation of  \BBz\ experiments is to cover the inverted-hierarchy region of Majorana neutrino mass. This region is indicated by other neutrino mass experiments to be between 15 and 50 meV. The various proposals to reach this goal are summarized in Table~\ref{tab:FutureExperiments}.

For a given experiment, the sensitivity to \mee\ can be written as~\cite{Moe91a}:
\begin{eqnarray}
\label{eqn:ExptLimit}
\mbox{\mee} &<& \frac{(2.50 \times 10^{-5} \mbox{meV})}{\mbox{\Mz}} \sqrt{\frac{N W}{M T fx\epsilon \mbox{\Gz}}}~,
\end{eqnarray}
where $M$ is the detector mass in kg, $T$ is the live time in years, $N$ is the upper limit on the number of counts assigned to signal, $W$ is the molecular weight of the detector material, $f$ is the isotopic abundance of the \BBz\ isotope, $x$ is the number of \BBz\ atoms per molecule, $\epsilon$ is the detection efficiency, and \Gz\ is the phase space factor. The constant in the equation has units appropriate for \mee\ given in meV. In most publications, $N$ takes the form of:
\begin{eqnarray}
N & = & \sqrt{N_B}  =  \sqrt{b\Delta E MT}	\mbox{  , background limited}	\nonumber \\  
  & = & 1.18								\mbox{  , background free, 68\% CL} 
\end{eqnarray}
where $N_B$ is the number of background counts, $b$ is the background index or number of background counts per energy and exposure (counts/keV-kg-y), and $\Delta E$ is the resolution-determined energy window (referred to as the Region of Interest or ROI) at the endpoint over which the background is measured. The product $M T$ is the exposure, and for an experiment to reach \mee\ sensitivity near 15 meV, requires exposures near 10 t-y and background rates below 1 count/t-y in the region of interest.

The most critical feature of any \BBz\ experiment is the background level. The dominant background in these experiments to date arises from $\alpha$, $\beta$ and $\gamma$ emitting isotopes contained as contaminants within the experimental apparatus. The most significant of these contaminants are the isotopes of the U and Th natural decay chains. All materials have U/Th in trace levels. Typical values are ppm or ppb. If a detector material has Th at a level of 1 ppb, one ton of that material will have an activity of $10^8$ decays per year. For neutrino masses within the inverted-hierarchy region, we expect a \BBz\ decay rate of about 1/t-y or less. Purifying material to a level that will allow the observation of this low rate is quite a challenge. And even when the technology is available, quality assurance is complicated at this stringent level. At the time of this writing, {\sc Gerda} has achieved the lowest background in the \BBz\ ROI~\cite{Agostini2013a}, about 40 counts/(t-y). This is a remarkable achievement but still a factor of 40 above what is required to investigate the inverted hierarchy region.

The first direct measurement of \BBt\ used a time projection chamber~\cite{Elliott1987}. This
 was a fairly large apparatus ($\approx$m$^3$) for a modest amount
of source (13 g) and, therefore, the design doesn't scale easily to large source mass with very low backgrounds. To consider how one might design a large experiment, Eq.~(\ref{eqn:ExptLimit}) can be used to develop a set of criteria for an ideal \BBz\ experiment~\cite{ell02,Elliott2003}. Such criteria for an experiment to reach the 15 meV goal, include:

\begin{itemize}
 \item The experimental exposure must be large enough ($M T \approx$ 10 t-y). A large quantity of isotope is required and the duty cycle of the experiment must be high. High isotopic abundance in the sample is required, as is a high efficiency of detection.
 \item The background in the ROI must be low enough ($N_B < 1$ count/t-y).
 \item Good energy resolution is required to reduce the background rate within the ROI. It is also required to prevent the tail of the \BBt\ spectrum extending into the \BBz\ ROI. Furthermore, good resolution can help prove that a peak is at the expected energy and is therefore due to \BBz\ in the case of an observation.
 \item A small detector volume minimizes internal backgrounds, when they scale with the detector volume. This is most easily accomplished by an apparatus whose source is also the detector. However, a very large source may have some advantage due to self shielding, although such a configuration may also result in some inefficient use of isotopic material.
 \item Event reconstruction, providing kinematic data such as opening angle and individual electron energy, can aid in the elimination of backgrounds. This data might also help elucidate the physics if a statistical sample of \BBz\ events is observed.
 \item Good spatial resolution and timing information can help reject background processes.
 \item If an experiment was able to observe the $^{N-2}_{Z+2}A$ daughter (see Eq.~\ref{eqn:BB0nu}) in coincidence with the \BB\ decay energy, it would eliminate most potential backgrounds except \BBt.
 \item The cost of these next generation experiments will be substantial. Therefore, any experiment must be based on a demonstrated technology for the detection of \BB. That is, one must demonstrate that one can achieve the required background.
 \item The nuclear theory is better understood in some isotopes than others. However, arguments have been made~\cite{Robertson2013} that there is no strongly preferred isotope when all the isotope related factors are considered in Eq.~\ref{eqn:ExptLimit}.  
 \item A high value for the energy available to the emitted $\beta$ particles is desired (i.e. a high Q-value), as it places the ROI above many potential backgrounds.
\end{itemize}

No experiment, past or proposed, is able to optimize for all of these characteristics simultaneously. Each collaboration has chosen a design that emphasizes different aspects of this list. In particular, the requirements of a large mass and good energy resolution are frequently at odds. The best resolution experiments use high-purity solid-state detectors (e.g. \MJ, {\sc Gerda}, CUORE). The cost of these detectors impedes instrumenting a large volume. Large quantities of scintillator with dissolved isotope can achieve large masses, but the lack of energy resolution reduces discovery potential (e.g. KamLAND-Zen, SNO+). Table~\ref{tab:FutureExperiments} summarizes the ideas proposed for the future. As can be seen from the number of listings, there are a lot of ideas being pursued (too many to discuss each here, unfortunately).

\begin{table*}[htdp]
\caption{A summary list of the \BBz\ proposals and experiments. The Q-Value is the available energy for the decay as referenced in the text.}
\begin{center}\begin{tabular}{|c|c|c|c|}
\hline
\hline
Isotope			&   Q-Value			& Technique							& Collaborations			 \\
				&	(MeV)			&									&						\\
\hline

\nuc{48}{Ca}		&	4.274			& CaF$_2$ scintillating crystals		&CANDLES\cite{Ume08}, CARVEL\cite{zde05}	\\
\hline

 \nuc{82}{Se}		&	2.995			& $\begin{array}{c} \mbox{ZnSe scintillating bolometers} \\ \mbox{Thin foils and tracking} \end{array}$		& $\begin{array}{c} \mbox{LUCIFER\cite{Arnaboldi11}} \\ \mbox{SuperNEMO\cite{Barabash2012}} \end{array}$								\\
\hline

\nuc{76}{Ge}		&	2.039			& high purity Ge semiconductor detectors	& {\sc Gerda}\cite{Agostini2013a}, {\sc Majorana}\cite{Abgrall2013}		\\
\hline

\nuc{100}{Mo}	&	3.034			& $\begin{array}{c} \mbox{CaMoO}_4 \mbox{bolometers} \\ \mbox{Thin foils and tracking} \\ \mbox{ZnMoO}_4 \mbox{bolometers} \end{array}$
	& $\begin{array}{c} \mbox{AMoRE\cite{Lee11}} \\ \mbox{MOON\cite{eji00}} \\ \mbox{Mo Bolometer\cite{Beeman2012}}\end{array}$								\\
\hline
	
\nuc{116}{Cd}		&	2.809		& CZT semiconductor detectors	& COBRA\cite{Dawson2009}		\\
\hline

\nuc{130}{Te}		&	2.528		& $\begin{array}{c} \mbox{TeO}_2 \mbox{bolometers} \\ \mbox{Te disolved in scintillator} \end{array}$
& $\begin{array}{c} \mbox{CUORE\cite{Alessandria2011}} \\ \mbox{SNO+\cite{Hartnell2012}} \end{array}$	\\
\hline

\nuc{136}{Xe}	& 2.458	& $\begin{array}{c} 	\mbox{liquid Xe time projection chamber} \\ 
													\mbox{Gaseous Xe time projection chamber} \\
													\mbox{Xe dissolved in scintillator} \\
													\mbox{Scint. liq. Xe within Graphene sphere} \end{array}$		&
								$\begin{array}{c} 	\mbox{EXO-200\cite{Auger2012}, nEXO}, \mbox{LZ\cite{Akerib2013a}} \\ 
													\mbox{NEXT\cite{Gom11}} \\
													\mbox{KamLAND-Zen\cite{Gando2013}} \\
													\mbox{GraXe\cite{GomezCadenas2012}} \end{array}$		\\
\hline
													
\nuc{150}{Nd}	&	3.371	& \mbox{thin foils and tracking}	& \mbox{DCBA\cite{ish00}}		\\
\hline

\nuc{160}{Gd}	& 	1.730 	& Cd$_2$SiO$_5$:Ce scint. crystals in liq. scint.	& GSO\cite{wan02}		\\			
\hline

Various			& 		& Quantum dots in liquid scintillator		& Quantum Dots\cite{Winslow2012,Aberle2013}		\\							
\hline
\hline
\end{tabular}
\end{center}
\label{tab:FutureExperiments}
\end{table*}%

\subsubsection{Proton Decay}
\label{sec:ProtonDecay}
There is a close connection between proton decay and double beta decay due to the relationship between $B$ and $L$ violation in extensions to the Standard Model. Using an argument along similar lines to the Schechter-Valle theorem, Babu and Mohapatra~\cite{Babu2014} have shown that if two $B$ violating decays are observed, with at least one that obeys the selection rule $\Delta(B-L) = \pm 2$, one will have established that neutrinos are Majorana.

\subsubsection{Accelerator Searches}
\label{sec:AccerSearch}
Processes that violate $L$ by 2 units are indicative of the exchange of a virtual Majorana neutrino. When the Majorana neutrino mass is light compared to the energy scale of the process, then the rate of that process will scale as the effective Majorana mass squared ($\langle m \rangle \approx |\sum_{light}U_{ei}^2m_i|$). This was the possibility discussed in detail for \BBz\ above and given in Eq.~(\ref{eqn:mee}). In contrast, if the mass is heavy compared to the energy scale, then the rate will scale inversely to the effective Majorana mass ($\langle M \rangle \approx |\sum_{heavy}\frac{V_{ei}^2}{M_i}|$). (Here we have used $V_{ei}$ in contrast to the previous $U_{ei}$ to emphasize that the mixing is among the heavy neutrino states.) Such heavy Majorana neutrinos, if they exist, would not only mediate \BBz, but also processes in high energy collisions at accelerators. In this later case, accelerators can compete with \BBz\ in the search for Majorana fermions under certain conditions.

In Table~\ref{tab:OtherDeltaL}, we list frequently searched-for processes that would indicate the existence of Majorana particles. These include: $\mu$ to e conversion, meson decay to two leptons, neutrino-antineutrino oscillation, di-lepton production and inverse double beta decay. Neutrinos are not observed in the detectors used to study these processes. Hence they not only carry away missing energy but also lepton number, which can make evidence for a Majorana particle difficult to confirm. Therefore the best limits come from processes that have lepton number violation but no missing energy. 

Atre, Han, Pascoli and Zhang wrote a nice review of the searches for heavy Majorana neutrinos~\cite{Atre2009}. A good overview of the field of lepton number conservation including the $\Delta L = 2 $ class that is indicative of Majorana neutrinos can be found in the work by de Gouvea and Vogel~\cite{DeGouvea2013}. An estimate of the rate of inverse double beta decay $e^-e^- \rightarrow W^-W^-$ is discussed by Rodejohann~\cite{Rodejohann2011} with a clear discussion of the issue of resonance. This latter reaction is the fundamental $\Delta L = 2$ process. 

For heavy Majorana neutrino masses below 5 GeV, meson decays produce good limits. The best are due to $K^+ \rightarrow l^+l^+\pi^-$~\cite{Atre2009} but only below 350 MeV. For the energies up to 5 GeV, one relies on charm decays from BaBar~\cite{Lees2011} and $B$ decays from BaBar~\cite{Lees2012}, CLEO~\cite{Edwards2002}, and LHCb~\cite{Aaij2012}. Limits on $\tau$ decays that provide constraints near 1 GeV were found by Belle~\cite{Miyazaki2013}. From 10-100 GeV, past results come from dilepton production in hadron collisions from CDF~\cite{Abulencia2007}, DELPHI~\cite{Abreu1997}, and L3~\cite{Adriani1992}. The most sensitive $\mu^- \rightarrow e^+$ conversion result comes from the SINDRUM II experiment~\cite{Kaulard1998}. None of these results compete with \BBz\ particularly well in constraining the hypothesis of Majorana neutrino exchange. 

The hadron collision search is well underway at the LHC. If $m_R$ is near the W boson mass ($M_W$), then the LHC may have sensitivity to Majorana neutrinos in dilepton production~\cite{Keung1983}, even given the constraints imposed by \BBz~\cite{Atre2009}. In general, the LHC program will expand the results in energy up to about 500 GeV. The lepton collision technique would be a powerful study from any future linear collider. The reaction $WW \rightarrow ll$ (See Fig.~\ref{fig:GenericFeynmann}) will have a resonant enhancement of its rate when the momentum transfer between the two particles is near that of $m_R$. 

\begin{table}[htdp]
\caption{A summary $\Delta L = 2$ processes that are studied to search for Majorana neutrinos. $L_0$ ($L_F$) is the initial (Final) $L$ in the reaction. These example reactions are just one of many possibilities for each row. For a more detailed discussion, experimental limits  rate estimates, and many more such examples, see Refs.~\cite{Atre2009,Rodejohann2011,DeGouvea2013}.}
\begin{center}\begin{tabular}{|c|c|c|c|}
\hline
\hline
Process		&   $L_{0}$		& Fin. $L_{F}$		& Example			 \\
\hline
Decay		& 0					& 2				& \BBz, $K^- \rightarrow \mu^-\mu^-\pi^+$ \\
Conversion	& $\pm$1				& $\mp$1		& $\mu^- + (Z,A) \rightarrow e^+ + (Z-2,A) $ \\
Lepton Decay	& $\pm$1				& $\mp$1		& $\tau^- \rightarrow e^+\pi^-\pi^-$ 		\\
Lepton Coll. & 2			& 0				& $e^-e^- \rightarrow W^-W^-$	\\
Hadron Coll. & 0			& -2				& $pp \rightarrow \mu^+\mu^+X $\\
						
\hline
\hline
\end{tabular}
\end{center}
\label{tab:OtherDeltaL}
\end{table}%

\subsubsection{Searches for a Fourth Neutrino}
There are 3 known neutrino mass eigenstates. Some tantalizing, but as yet not universally accepted, evidence for a fourth neutrino comes from the LSND~\cite{agu01} and MiniBooNE~\cite{Aguilar2009,Aguilar2010} experiments. These experiments are accelerator neutrino oscillation experiments that find their data best described by a neutrino mass difference that does not match that of either the solar or atmospheric oscillation results. Hence, if those results are established, it implies an additional light neutrino state. 

When we discussed the seesaw mechanism in Sec.~\ref{sec:seesaw}, we focused on the assumption that $m_R$ is very large compared to $m_L$ and $m_D$. However, if $m_D \gg m_L,m_R$ or if $m_D \approx m_L$ and/or $m_R$, then the spectrum of neutrino masses required by the LSND and MiniBooNE results can be accommodated. Hence, the discovery of a fourth neutrino mixing with the 3 known neutrinos would not prove that neutrinos are Majorana, but it does fit that paradigm well.

There are a large number of searches, ongoing and proposed, for a fourth neutrino mass eigenstate. These searches include experiments using neutrinos produced by accelerators, reactors, and intense radioactive sources, as well as observations in astrophysics and cosmology. A detailed summary of these searches, in addition to an overview of the theory, is given by Abazajian {\em et al.}~\cite{Abazajian2012}. That review provides summary tables such as those we have provided here for the other classes of searches. Therefore, we do not provide one ourselves.

\subsubsection{Dark Matter Searches}
\label{sec:DMSearch}
Dark matter comprises a large fraction of the Universe's energy density and a dominant fraction of the Universe's matter. The nature of dark matter is not fully understood and a large program is underway to elucidate it. The recent long-range planning process for high energy physics in the US has resulted in a number of nice reviews~\cite{Buckley2013,Cushman2013,Kusenko2013} detailing this program, and the review by Schumann~\cite{Schumann2013} ties the field together well in a short summary. The review by~\textcite{Jungman1996} gives details about SUSY and its relationship to dark matter. We refer the interested reader to these reviews and only briefly summarize the situation here. 

There are many candidates for the dark matter particle, which we will indicate by $\chi_0$, and many of these candidates are Majorana in nature. Detecting dark matter will not, by itself, be a smoking gun that Majorana particles exist, but it would certainly be a key piece of data addressing the question. To accommodate the astrophysical and cosmological data, these particles must be electrically neutral, non-relativistic and stable.  There are 3 techniques used to search for these particles, including direct detection, indirect detection, and colliding beam experiments. 

Direct detection searches hope to observe Weakly Interacting Massive Particles (WIMP) by their interaction within a low-background particle detector located deep underground. Supersymmetry allows the existence of a particle whose mass and cross section match well with that required to explain the relic density of dark matter. Furthermore, it would be a Majorana particle. The signature for the resulting reaction
\begin{equation}
\chi_0 + ^N_ZA  \rightarrow  \chi_0 + ^N_ZA ,
\end{equation}
is the recoiling nucleus, which is detected. The recoil energy, however, is low, depositing a few to ten's of keV. Furthermore, the energy spectrum of the recoil is nondescript, being exponentially decreasing. The expected event rates are also extremely low. Hence, the detector requirements are stringent with respect to the low energy threshold and background. Seeing a significant signal above background with different targets can provide a consistency test that dark matter is truly being observed. Presently, the most sensitive detectors are those based on liquid noble gas targets. The LUX experiment~\cite{Akerib2013}, sited in the Sanford Underground Research Facility, has the best limits on the WIMP detection rate at the time of this writing. 

Indirect detection searches look for processes within astrophysical objects such as
\begin{equation}
\chi_0 + \bar{\chi}_0 \rightarrow q + \bar{q},~~ l + \bar{l},~~ W^+ + W^-,~~ \mbox{or}~~Z + Z ,  
\end{equation}
where $q$ represents a quark, $W$ represents a W-boson, and $Z$ represents a Z-Boson. These product particles then decay into particles that can be observed. Heavy astrophysical bodies such as the Sun, dwarf galaxies or the galactic center can gather $\chi_0$'s into a locally high density that enhances the annihilation rate. Satellites looking for anti-particles and experiments searching for high-energy $\gamma$ rays in the cosmic ray spectrum place limits on this annihilation. The detection scheme for these experiments assumes the particle self annihilates, which is a key characteristic of Majorana particles, such as the neutralino.

The cosmic microwave background multipole spectrum would be modified by dark matter annihilation in the early Universe. The lack of an observed impact on the multipole spectrum leads to significant constraints, especially for low-mass ($\approx$10 GeV) WIMP annihilation into electron-positron pairs~\cite{Galli2011}.

{\bf Collider beam} searches look for processes such as
\begin{equation}
q + \bar{q} \rightarrow \chi_0 + \bar{\chi}_0 + X ,
\end{equation}
where the $X$ represents a radiated $\gamma$, $Z$, $W$ or gluon. The $\chi_0$, being weakly interacting, is not observed within the detector; however, $X$ produces a monojet which is observed. Therefore, the signature is a large missing energy event with a monojet.

\section{Majorana Zero Modes in Solid State Systems}
\label{Sec:MajFermSolidState}
\label{Sec:SolidState}

Majorana fermions have been of great interest in condensed matter physics over the past decade.
In this review we shall focus on Majorana zero modes (MZMs) in solid state systems where they arise as emergent quasiparticles of the underlying
superconducting state. Although there are other proposed realizations,
superconducting systems are conceptually simplest, best understood,
and arguably closest to unambiguous physical realization and
detection. There exist several excellent reviews of this topic aimed at the
condensed matter audience \cite{Alicea2012,Beenakker2012,Tewari2013}. By contrast, our objective here  is to present the topic to a
broader audience of physicists who possess basic understanding of
condensed matter physics but are not specialists.


\subsection{Majorana Zero Modes}

As we already explained in Sec. II.B, quasiparticle excitations in
superconductors posses all the key attributes of Majorana fermions.
A situation of particular interest arises when the spectrum of
excitations in a superconductor is such that there
exists a single mode with exactly zero energy,
\begin{equation}\label{bdg2}
 H_{\rm BdG}(\br)\Phi_0(\br) = 0,
\end{equation}
separated from all other modes by an energy gap. 
According to the discussion presented in Sec. II.B we must conclude that, remarkably, only one
half of that mode is actually physical.
In addition, such a zero mode is self-conjugate under the symmetry defined in
Eq.\ (\ref{ph0}) meaning that 
\begin{equation}\label{ph2}
\Phi_0(\br)=\tau^y \sigma^y \Phi^*_0(\br),
\end{equation}
or, written in terms of the individual components, 
\begin{equation}\label{ph3}
\begin{pmatrix}
u_{0\uparrow}\\ u_{0\downarrow} \\ 
v_{0\uparrow} \\ v_{0\downarrow} 
\end{pmatrix}=
\begin{pmatrix}
-v^*_{0\downarrow}\\ v^*_{0\uparrow} \\ 
u^*_{0\downarrow} \\ -u^*_{0\uparrow} 
\end{pmatrix},
\end{equation}
where we have dropped the position argument for the sake of
clarity. The zero-mode annihilation operator is now given by Eq.\
(\ref{psi1}) which, when expanded, reads 
\begin{eqnarray}\label{psi2}
\hat\psi_0=i\int d^dr \biggl[
u^*_{0\uparrow}(\br)c_{\br\uparrow}&+&
u^*_{0\downarrow}(\br)c_{\br\downarrow}\\
&-& 
v^*_{0\uparrow}(\br)c^\dagger_{\br\downarrow} +
v^*_{0\downarrow}(\br)c^\dagger_{\br\uparrow}
\biggr],\nonumber
\end{eqnarray}
where the arbitrary $i$ factor has been added for convenience.
With help of Eq.\ (\ref{ph3}) it is now straightforward to verify that 
\begin{equation}\label{psi3}
\hat\psi^\dagger_0=\hat\psi_0,
\end{equation}
informing us that the zero mode particle is the same as the antiparticle and  is therefore Majorana.

To motivate interest in Majorana zero modes we note the following points:

(i) Because according to Eq.\ (\ref{bdg2}) it costs zero energy to create the particle described by $\hat\psi^\dagger_0$ we conclude that in the presence of the MZM the ground state of the system is degenerate: if $|0\rangle$ is a ground state then so is $\hat\psi^\dagger_0|0\rangle$. However, because one cannot form the usual number operator from Majorana fermions $(\hat\psi^\dagger_0\hat\psi_0=\hat\psi_0\hat\psi_0=1)$ it is not possible to label the degenerate ground states by the number of MZMs; technically the degeneracy in the presence of a single MZM is $\sqrt{2}$. We shall elucidate this mysterious statement below.

(ii) The discussion above makes it clear
that if a Majorana zero mode exists in a system then it is
topologically protected, provided that there is an energy gap (often
called a `minigap') separating it from all other states. The reason is
that the zero mode
cannot acquire a non-zero energy $E_0$ by any continuous deformation
of the Hamiltonian that does not close the minigap. If this were so then the 
symmetry  defined in Eq.\ (\ref{ph0}) would require another mode to
appear at energy $-E_0$, in violation of the
unitary evolution.

(iii) A single unpaired MZM can exist only in an infinite system. In  systems of finite size
Majorana modes always appear in pairs, reflecting the fact that such
systems always contain an integral number of electrons. Nevertheless,
a situation of interest arises when the two MZMs are
spatially separated so that their individual wavefunctions have a
negligible overlap. In this case, when probed locally, the system
exhibits an unpaired MZM. Also, in this situation
Majorana modes can be moved away from zero energy without closing the minigap by simply
bringing  them close together so that the wavefunctions overlap. 
The two zero modes thus evolve into a pair of levels $(E_0,-E_0)$ with
a splitting proportional to the overlap.

(iv) If there are several MZMs in the system it is
  easy to show that their associated annihilation operators
  $\hat\psi_{0j}$ formally satisfy the canonical commutation relations
  (\ref{can2}) characteristic of Majorana fermions. Upon closer
  examination, however, it turns out that their exchange statistics is
  more complicated and the zero modes should be correctly described as
  non-Abelian anyons \cite{Moore1991, Nayak2008}. This very interesting  property arises because
  of the extra contribution to the exchange statistics coming from the
  superconducting condensate and we will discuss it in more detail
  later in this Section. Non-Abelian exchange statistics also
  underlies the potential significance of MZMs for
  topological quantum computation.

Unpaired MZMs discussed thus far are localized objects
that are typically associated with point-like defects present in
topological superconductors. We will explain in the following
subsections how they arise at the ends of 1D SC wires, in the vortex
cores of 2D topological superconductors and other situations. We will
also discuss their unusual properties that underlie much of the current
interest in these exotic forms of quantum matter.

\subsection{Kitaev Chain: the 1D Prototype System}
\label{sec:kitaev}
The simplest model system that shows unpaired MZMs is the
Kitaev chain \cite{Kitaev2001}. It describes a 1D system of {\em
  spinless} fermions and for this reason it has been initially viewed as a
somewhat unphysical toy model. However, it has been realized more
recently that in the presence of spin-orbit coupling and the Zeeman
field real electrons can in fact behave essentially like spinless
fermions. Kitaev model, which is simple and exactly soluble, thus
provides an extremely useful paradigm for MZMs in one
spatial dimension and for this reason we shall discuss it in some
detail.

The Kitaev chain consists of spinless fermions hopping between the
sites of a 1D lattice described by the  Hamiltonian 
\begin{eqnarray}\label{hkit1}
{\cal H}=\sum_j\biggl[-t (c^\dagger_j c_{j+1}+{\rm h.c.}) &-& \mu  (c^\dagger_j c_{j}-{1\over 2}) \\
&+& (\Delta c^\dagger_j c^\dagger_{j+1}+{\rm h.c.})\biggr],\nonumber
\end{eqnarray}
where $\Delta$ represents the nearest-neighbor pairing amplitude, the
simplest allowed possibility for SC order with spinless fermions. Henceforth we shall assume for the sake of simplicity that $\Delta$ is real and consider a chain with $N$ sites and open boundary conditions. Using transformation (\ref{can3}) we can rewrite this Hamiltonian in the Majorana representation,
\begin{eqnarray}\label{hkit2}
{\cal H}={i\over 2}\sum_j\biggl[-\mu \gamma_{j,1}\gamma_{j,2} &+&(t+\Delta)\gamma_{j,2}\gamma_{j+1,1} \\
&+&(-t+\Delta)\gamma_{j,1}\gamma_{j+1,2}\biggr]. \nonumber
\end{eqnarray}

\begin{figure}[t]
\includegraphics[width = 7.5cm]{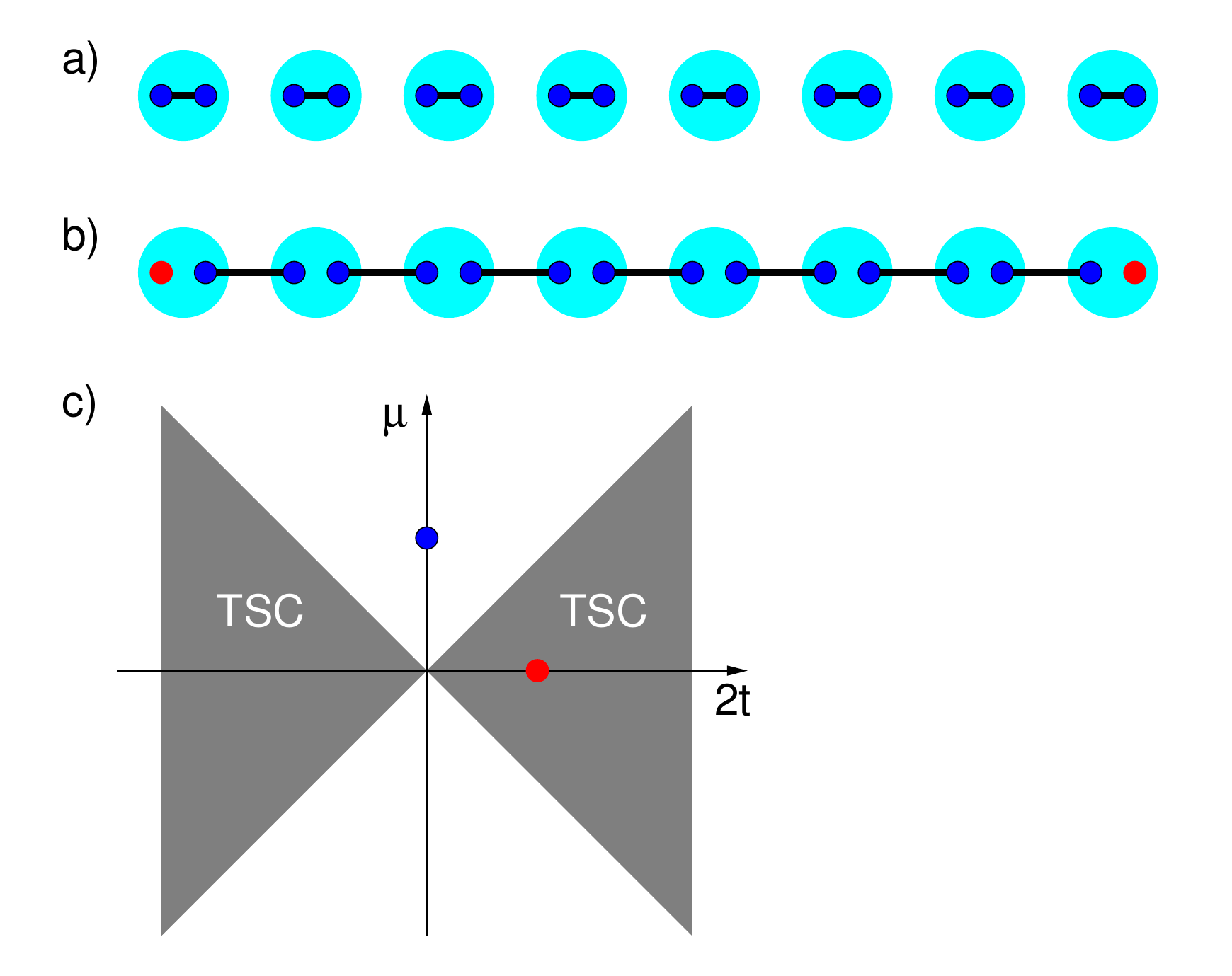}
\caption{Two phases of the Kitaev chain. a) In the trivial phase
  Majorana fermions on each lattice site can be thought of as bound
  into ordinary fermions.  b)  In the topological phase Majoranas on
  neighboring sites are bound leaving two unpaired Majorana fermions
  at the ends of the chain. c) The phase diagram of the Kitaev chain
  in the $\mu$--$2t$ plane, showing the topological phase (TSC) and
  the normal phase. The dots mark the special points
  in the parameter space considered in the text.     
}\label{figss2}
\end{figure}

At this point it is useful to examine two specific limits. First, consider the case $\Delta=t=0$. The Hamiltonian becomes simply
\begin{equation}\label{hkit3}
{\cal H}={i\over 2}(-\mu)\sum_j \gamma_{j,1}\gamma_{j,2}=-\mu\sum_j(c^\dagger_j c_{j}-{1\over 2}).
\end{equation}
The ground state consists of all fermion states at site $j$ either occupied ($\mu>0$) or empty ($\mu<0$) and this is clearly a topologically trivial phase. Second, consider the case $\Delta=t$ and $\mu=0$. Now the Hamiltonian takes the form
\begin{equation}\label{hkit4}
{\cal H}=it\sum_{j=1}^{N-1} \gamma_{j,2}\gamma_{j+1,1}.
\end{equation}
The ground state of this Hamiltonian is easily found by defining a new set of fermionic operators
\begin{equation}\label{can11}
a_j={1\over 2}(\gamma_{j,2}+i\gamma_{j+1,1}), \ \ \ a_j^\dagger={1\over 2}(\gamma_{j,2}-i\gamma_{j+1,1}),
\end{equation} 
for $j=1, 2 ... N-1$. These live on nearest neighbor bonds of our 1D
chain as illustrated in Fig.\ \ref{figss2}b. In terms of these new fermions we have 
\begin{equation}\label{hkit3}
{\cal H}=2t\sum_{j=1}^{N-1}(a^\dagger_j a_{j}-{1\over 2}),
\end{equation}
and the ground state for $t>0$ is simply an $a_j$ vacuum with total energy $E_g=-t(L-1)$. The remarkable thing is that Hamiltonian (\ref{hkit4}) does not contain operators $\gamma_{1,1}$ and $\gamma_{N,2}$. These represent zero-energy MZMs localized at the ends of the chain. Together they encode one Dirac fermion which is fundamentally delocalized between the two ends of the chain. We remark that similar considerations yield unpaired MZMs also for the special case $\Delta=-t$ and $\mu=0$.   

The two special cases considered above represent two distinct phases of the Kitaev model: the trivial phase and the topological phase with unpaired MZMs localized at its ends. To show that these indeed correspond to stable phases consider the same Hamiltonian (\ref{hkit1}) but now with periodic boundary conditions. In momentum space it can be written as     
\begin{equation}\label{hkit5}
{\cal H}=\sum_q\left[(-2t\cos{q}-\mu)c^\dagger_qc_q+\Delta(i\sin{q}\ c_qc_{-q}+{\rm h.c.}\right],
\end{equation}
and has an excitation spectrum of the form
\begin{equation}\label{ex5}
E(q)=\pm\sqrt{(2t\cos{q}+\mu)^2+(\Delta\sin{q})^2}.
\end{equation}
If we now focus on the superconducting phases (i.e.\ $\Delta\neq 0$)
then it is easy to see that the excitation spectrum Eq.\ (\ref{ex5})
remains fully gapped except when $2t=\pm\mu$. This condition defines
two lines, indicated in Fig.\ \ref{figss2}, which mark the phase
boundaries between the two stable phases of the model. (We are making
use of the general principle of adiabatic continuity which states that
two gapped phases are identical if they can be smoothly deformed into
one another without closing the excitation gap.) We identify the region $|2t|>|\mu|$ as the topological phase since the second special point considered above lies within this phase. The other phase is topologically trivial. 

Since the two phases have the same physical symmetries the transition
between them is a special type of a phase transition called
topological phase transition. The two phases are distinguished by the
presence or absence of unpaired MZMs at the ends in the
geometry with open boundary conditions. The question that naturally
arises and that will have important consequences in our search for
topological phases in realistic systems is the following. In the
absence of symmetry distinction is it possible to theoretically distinguish the
topological phase from the trivial phase by studying the bulk of the
system?  The answer is affirmative: such phases can be distinguished
by means of {\em topological invariants}. Among the better known
topological invariants are the Chern number, allowing one to
differentiate between different quantum Hall phases in two-dimensional
quantum Hall systems, and the more recently discovered $Z_2$ invariants
that characterize topological insulators in two and three
dimensions. For 1D topological superconductors the relevant
topological invariant is the Majorana number ${\cal M}=\pm 1$ first
formulated by Kitaev \cite{Kitaev2001}. In his seminal 2001 paper
Kitaev showed that all 1D fermionic systems with SC order fall into
two categories distinguished by  ${\cal M}$. Presence of unpaired
MZMs is indicated when ${\cal M}=-1$ and the system is gapped.

As topological invariants go ${\cal M}$ is easy to evaluate (although
the reasoning behind its formulation is more involved and we refer
the interested reader to the original Kitaev paper).  A Hamiltonian for any non-interacting  translationally invariant fermionic system in 1D can be written in the Majorana representation as 
\begin{equation}\label{hkit7}
{\cal H}={i\over 4}\sum_{lm\alpha\beta}B_{\alpha\beta}(l-m)\gamma_{l\alpha}\gamma_{m\beta},
\end{equation}
where $l$, $m$ denote the lattice sites while $\alpha$, $\beta$ label all
other quantum numbers, including spin and orbital degrees of
freedom. The Majorana number is defined as 
\begin{equation}\label{maj1}
{\cal M}={\rm sgn}\left\{ {\rm Pf}[\tilde B(0)] {\rm Pf}[\tilde B(\pi)]\right\}, 
\end{equation}
where $\tilde B(q)$ denotes the spatial Fourier
transform of $B(l-m)$  viewed as a matrix in indices $\alpha$,
$\beta$ and Pf$[A]$ denotes the Pfaffian (the square root of
determinant with a definite sign). For
$q=0,\pi$ matrix $\tilde B(q)$ is antisymmetric  (this follows from the requirement that ${\cal H}$ is
hermitian) and the Pfaffian is thus well defined.
For a known matrix of a small size the Pfaffian is readily evaluated.
The topological invariant for a 1D superconductor can be
therefore easily computed directly from the system's Hamiltonian. We will present a concrete example of such
a computation below. 

An important conceptual tool follows from studying
the limit of weak SC order, i.e.\ the situation when $\Delta$ is much
smaller than all relevant energy scales in the problem, such as the bandwidth, which is
often the case in superconductors. In this limit one can show
that Eq.\ (\ref{maj1}) further simplifies to
\begin{equation}\label{maj2}
{\cal M}=(-1)^\nu,
\end{equation}
where $\nu$ represents the number of Fermi points $q_F$  of the underlying
normal system $(\Delta=0)$ in the interval $(0,\pi)$. This formulation
provides a simple but extremely useful guide to searches for
topological superconductors in 1D: one is compelled to look for 1D
metals with an odd number of Fermi points in the right half of the
Brillouin zone. Such 1D metals, when made superconducting, form a
topological phase with unpaired MZMs localized
at their ends.  We emphasize that the classification implied by Eqs. (\ref{maj1}) and (\ref{maj2}) is physically meaningful only when applied to fully gapped systems. Some concrete examples of physical systems where such
a situation can occur are given in the following subsection. A more
comprehensive discussion is given in the existing review articles
\cite{Alicea2012,Beenakker2012,Tewari2013}.  

We close this subsection by evaluating ${\cal M}$ for the Kitaev model
using Eq.\ (\ref{maj1}). To this end passing into momentum space  we can write the Kitaev Hamiltonian (\ref{hkit2}) in the following form,
\begin{equation}\label{hkit8}
{\cal H}={i\over 4}\sum_q(\gamma_{q1},\gamma_{q2})
\begin{pmatrix}
0 & D_q \\
-D^*_q & 0 \\
\end{pmatrix}
\begin{pmatrix}
\gamma_{-q1} \\
\gamma_{-q2}
\end{pmatrix},
\end{equation}
with $D_q=-\mu-2t\cos{q}-2i\Delta\sin{q}$. Pfaffian of a $2\times 2$ antisymmetric matrix is simply given by its upper off-diagonal component which yields the Majorana number 
\begin{equation}\label{maj3}
{\cal M}={\rm sgn}(D_0 D_\pi)={\rm sgn}(\mu^2-4t^2). 
\end{equation}
The topological phase occurs when ${\cal M}=-1$, or $|\mu|<2|t|$, in
accord with our earlier analysis. One can also test Eq.\
(\ref{maj2}). When $\Delta=0$ the normal state dispersion of the
Kitaev chain becomes $\epsilon(q)=-2t\cos{q}-\mu$. This yields one
Fermi point between $(0,\pi)$ when $|\mu|<2|t|$ and no Fermi points
otherwise, confirming once again the structure of the topological
phase diagram indicated in Fig.\ \ref{figss2}c.

\subsection{Physical Realizations in 1D}
The key obstacle standing in the way of physical realizations of
Kitaev's paradigm is the electron spin. In most natural realizations of a 1D chain, electron spin causes all bands to be doubly degenerate thus preventing the desirable situation with an odd number of Fermi points, required for the emergence of the topological phase according to the discussion in the previous subsection. Below we discuss special situations in which this problem can be avoided. They involve various combinations of the spin-orbit coupling and magnetic interactions that produce a normal metal that is effectively spinless. Superconductivity is then induced through the proximity effect, whereby pairing occurs due to Cooper pair tunneling from a nearby superconductor.

\subsubsection{Edge of a 2D Topological Insulator}    
\label{sec:2DTI}
Topological insulators (TIs) are materials with gapped insulating bulk and topologically protected gapless surface states \cite{Moore1, Kane1, Franz2013, Qi2011}. The surface states form an unconventional metal and are protected by time reversal symmetry $\cT$. This remarkable behavior comes about as a result of strong spin-orbit coupling and occurs in crystals and alloys made of heavy non-magnetic elements. Canonical examples of TIs include HgTe quantum wells, Bi$_x$Sb$_{1-x}$ alloys and Bi$_2$Se$_3$ crystals.

\begin{figure}[t]
\includegraphics[width = 8.5cm]{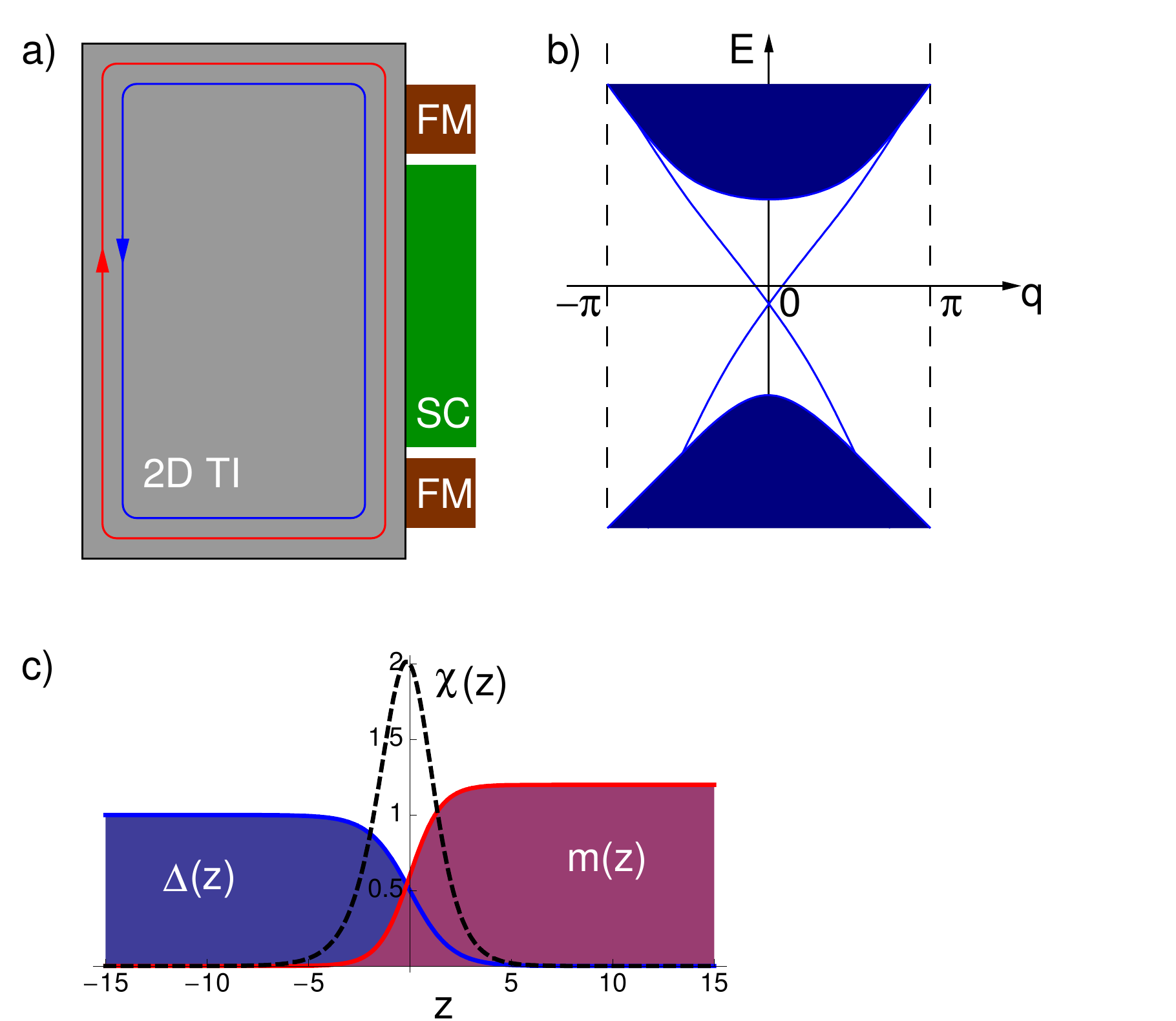}
\caption{a) 2D topological insulator interfaced with a superconductor (SC)
  and two ferromagnetic insulators (FM). MZMs are expected
  occur at  the SC/FM boundaries. b) Schematic spectrum of the surface
  state in a 2D TI. The shaded regions represent the bulk conduction
  and valence bands. c) SC and magnetic order parameter profiles near
  the SC/FM boundary assumed in the calculation. The dashed line shows
  the resulting Majorana wavefunction amplitude.
}\label{figss3}
\end{figure}

The edge of a 2D TI is characterized by a pair of counter-propagating,
spin filtered, linearly dispersing edge states, illustrated in
Fig.\ \ref{figss3}b. The relevance of such an edge for the emergence of MZMs stems from the fact that when the chemical potential resides
inside the bulk bandgap, the state can be viewed as a 1D system with an
odd number of Fermi points in the right half of the Brillouin
zone. Thus, according to the Kitaev criterion, one expects such an
edge to form a foundation for a 1D topological superconductor if
superconductivity can be induced, e.g. by the proximity effect. One
important subtlety stems from the fact that being a boundary of a 2D
system, the edge does not have an end. In order to localize the
expected MZMs one must employ a setup illustrated in
Fig.\ \ref{figss3}a with a TI edge interfaced with a superconductor
and a magnetic insulator \cite{Kane2}. The magnetic material also
provides the $\cT$-breaking that is necessary to obtain unpaired
MZMs. As we shall show momentarily, MZMs 
arise at the boundary between the SC and magnetic regions of the edge. 

The low-energy theory of the edge modes is described by the Bloch Hamiltonian
\begin{equation}\label{h2dti1}
h_0(q)=vq\sigma^y+m\sigma^x -\mu
\end{equation}
where $v$ is the mode velocity and $m$ represents the $x$-component of
the magnetic order induced by the proximate magnetic insulator. We are
selecting specific quantization directions in the spin space but the
only important feature is that the magnetic order has a component
perpendicular to the electron spin. In this case the spectrum of
excitations reads $\epsilon(q)=\pm\sqrt{v^2q^2+m^2}-\mu$ and is
illustrated in Fig.\ \ref{figss3}b. Magnetic order will gap out the surface mode if $|\mu|<|m|$. This explains why magnetic order can be thought of as effectively ending the wire. Since we shall be interested in spatially varying situations we now pass into real space taking $q\to-i\partial_z$ in Eq.\ (\ref{h2dti1}). To include SC order we then follow the same path that led us to Eq.\ (\ref{hbdg1}) and obtain the BdG Hamiltonian relevant to this situation
\begin{equation}\label{h2dti2}
H_{\rm BdG}=v(-i\partial_z)\sigma^y\tau^z+m(z)\sigma^x+\Delta(z)\tau^x,
\end{equation}
where we have set $\mu=0$ and assumed that $\Delta(z)$ is real. To
find the expected MZM we seek a zero-energy eigenstate
$H_{\rm BdG}\Phi_0(z)=0$ in the vicinity of a domain wall between SC
and magnetic domains, as illustrated in Fig.\ \ref{figss3}c. It is a simple matter to show that there exists precisely one such zero mode with the wavefunction
\begin{equation}\label{h2dti3}
\Phi_0(z)={\chi_0(z)\over \sqrt{2}}
\begin{pmatrix}
0 \\ 1 \\ -1 \\ 0
\end{pmatrix}, \ \ 
\chi_0(z)=Ae^{-{1\over v}\int_0^zdz'[m(z')-\Delta(z')]},
\end{equation}
where $A$ is a normalization constant. The corresponding field operator takes the form
\begin{equation}\label{h2dti4}
\hat\psi_0={1\over\sqrt{2}}\int dz\chi_0(z)[c_\downarrow(z)+c_\downarrow^\dagger(z)],
\end{equation}
and obeys the requisite Majorana condition $\hat\psi_0^\dagger=\hat\psi_0$. We note that the other domain wall (with SC domain to the right of the magnetic domain) also hosts an MZM which involves spin-up electrons. Also, we remark that MZMs persist for an arbitrary complex order parameter $\Delta$ and for $|\mu|<|m|$, although the solution is slightly more complicated when $\mu\neq 0$ with the wavefunction $\Phi_0(z)$ oscillating at the relevant Fermi wavevector.

\subsubsection{Nanowire Made from a 3D Topological Insulator }
\label{sec:TIwires}
The surface of a 3D topological insulator, such as Bi$_2$Se$_3$ and Bi$_2$Te$_3$, is known to host a single gapless linearly dispersing Dirac fermion. The massless character of this surface state is protected by $\cT$. For a planar surface perpendicular to the $z$ axis the low-energy theory of  the Dirac mode is described by the Bloch Hamiltonian \cite{Kane1}
\begin{equation}\label{h3dti1}
h_0(\bq)=v(q_x\sigma^y-q_y\sigma^x) -\mu.
\end{equation}
\begin{figure}[t]
\includegraphics[width = 8.5cm]{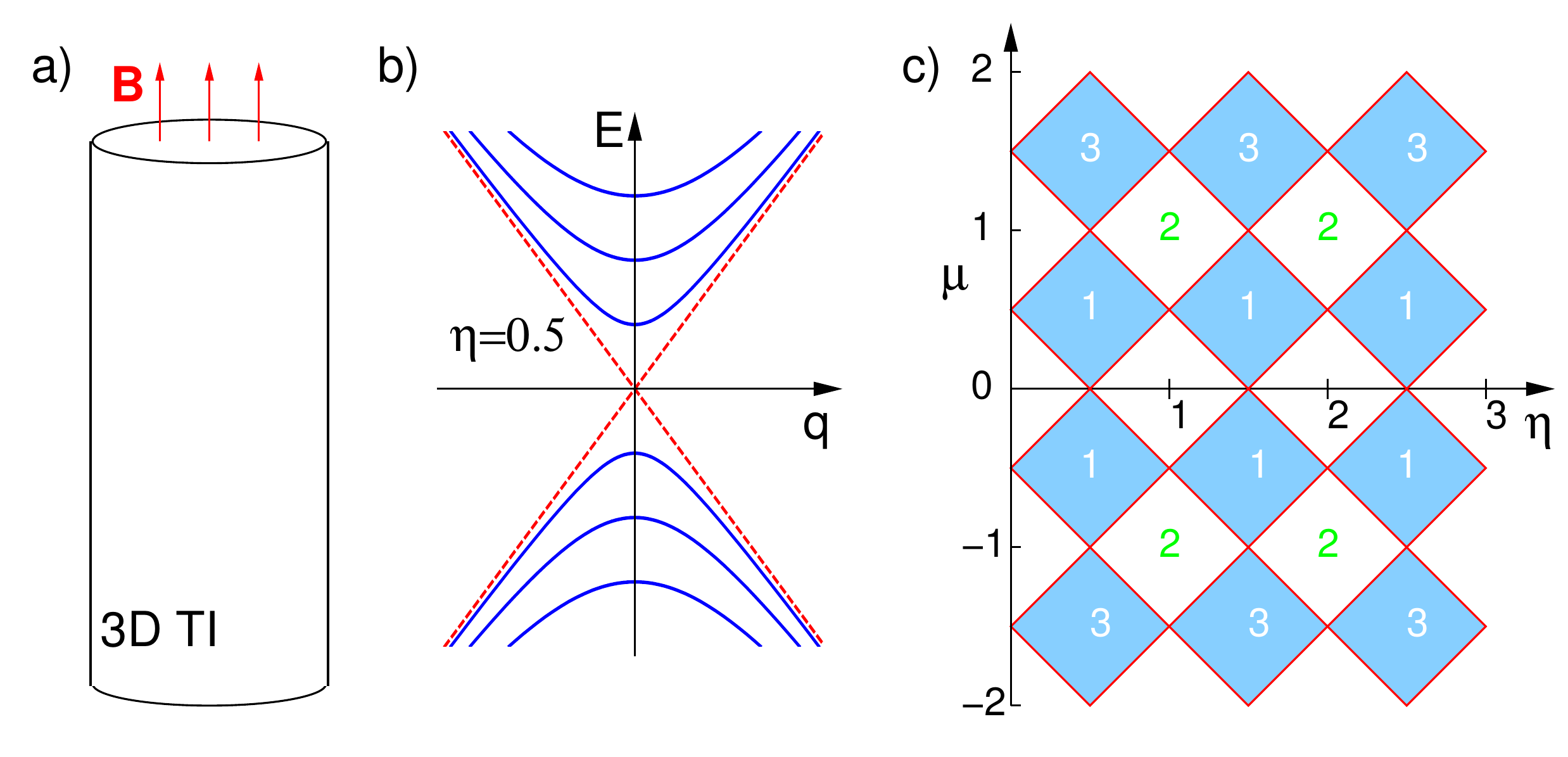}
\caption{a) 3D topological insulator wire in longitudinal magnetic
  field $\bB$. b) Spectrum of the surface
  state excitations when the total flux piercing the wire
  $\eta=\Phi/\Phi_0=SB/\Phi_0$ is half-integral. The dashed lines
  indicate non-degenerate bands while those represented by the solid
  lines are doubly degenerate.  c) The topological phase diagram in
  the limit of small $\Delta$. Shaded regions represent the
  topological phase and the numerals indicate the number of Fermi
  points in the right half of the Brillouin zone. 
}\label{figss4}
\end{figure}
We shall be interested in a quasi-1D wire geometry, illustrated in
Fig.\ \ref{figss4}a.
The wire exhibits Majorana end-modes when it is threaded by magnetic flux of half-integer flux quanta and brought into proximity of a superconductor \cite{cook2011}. To see this we note that  
a Dirac Hamiltonian analogous to Eq. (\ref{h3dti1}) can be formulated for an arbitrary curved surface \cite{mirlin1} and solved exactly for a surface of an infinitely long cylindrical wire with radius $R$ threaded by magnetic flux $\Phi$ \cite{rosenberg1}. The solution yields an excitation spectrum of the form
\begin{equation}\label{h3dti2}
\epsilon_l(q_z)=\pm v\hbar\sqrt{q_z^2+\left({l+\eta-1/2\over R}\right)^2},
\end{equation}
where $q_z$ is the momentum along the axis of the cylinder, $l=0,\pm
1,\dots$ is the angular momentum and $\eta=\Phi/\Phi_0$ is the
magnetic flux in units of flux quantum $\Phi_0=hc/e$. In the absence of the
magnetic field the system exhibits a finite-size gap $v\hbar/R$ and all levels are doubly degenerate. A more interesting situation arises when $\eta$ is half-integer (e.g. when $\Phi={1\over 2}\Phi_0$). In this case a single non-degenerate gapless
branch arises of the form $\pm v\hbar q_z$, while all other branches
remain doubly degenerate. This situation is illustrated in Fig.\
\ref{figss4}b. The spectrum now exhibits an odd number of Fermi points
in the right half of the Brillouin zone when the chemical potential
resides inside the bulk bandgap and we expect a topological state to
emerge when superconductivity is induced. The resulting topological
phase diagram is depicted in  Fig.\ \ref{figss4}c.

When the chemical potential is small, $|\mu|<v\hbar/2R$, so that it intersects only the lowest band, one can easily solve for the MZMs. The procedure is essentially identical to the one discussed above for the edge of the 2D TI and yields an MZM  at each end of the wire. When more bands are occupied and for wires with a non-circular cross section, such as those grown experimentally, the formula (\ref{h3dti2}) for the normal state spectrum holds only approximately but the pattern of degeneracies (i.e. one gapless non-degenerate branch plus a set of gapped doubly degenerate bands) remains robust as a consequence of $\cT$-invariance. One thus expects MZMs to exist in this case as well and this is indeed confirmed by numerical calculations \cite{cook2011,cook2012}.

\subsubsection{Semiconductor Quantum Wires}
\label{sec:smwires}
Another platform for MZMs is based on ordinary
semiconductors with strong spin-orbit coupling (SOC), as realized by  InSb or
InAs. This platform has gained a significant momentum recently due to
the existing expertise and technological background available for
these long studied materials. The initial proposal involved a 2D
structure composed of semiconductor and superconductor films
interfaced with an insulating ferromagnet \cite{Sau2010} or in the presence of
external in-plane magnetic field \cite{Alicea2010}. Later, advantages of 1D quantum
wires have been recognized \cite{Lutchyn2010,Oreg2010} and this is now the leading solid-state candidate for the experimental realization of unpaired MZMs. We shall focus on this system.

\begin{figure}[t]
\includegraphics[width = 8.5cm]{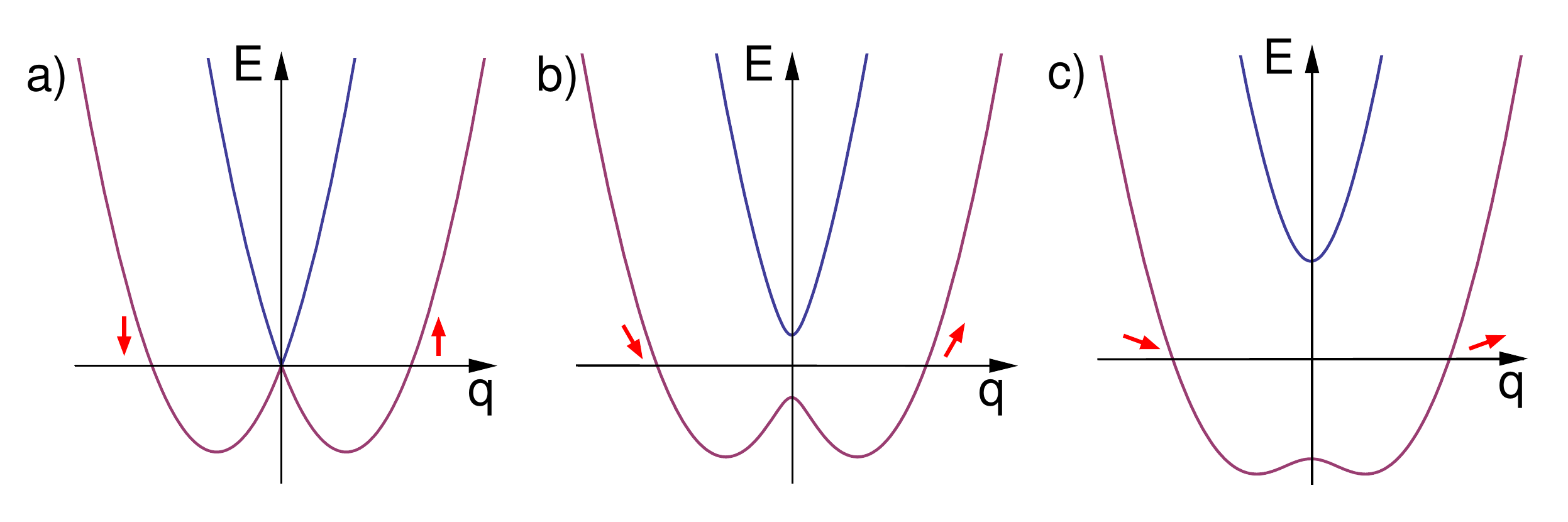}
\caption{ Excitation spectra of a single channel semiconductor quantum
  wire Eq.\ (\ref{hsemi2}). The three panels show representative
  cases with different Zeeman coupling, a)$V_Z/E_{\rm SO}=0.0$,
  b) 0.4 and c) 1.2. The arrows indicate the spin direction at the
  Fermi points. 
}\label{figss5}
\end{figure}

To understand the physics behind this proposal we start from a
Hamiltonian describing the low-energy electrons in a 1D quantum wire
with a Rashba SOC,
\begin{equation}\label{hsemi1}
{\cal H}(q_z)={\hbar^2 q_z^2\over 2m_{\rm eff}}+\alpha\hat{\bf n}\cdot(\bsig\times\bq).
\end{equation}
Here $m_{\rm eff}$ is the electron effective mass, $\alpha$ sets the
strength of the SOC, and  $\hat{\bf n}$ its direction. For a quantum
wire supported by a substrate, $\hat{\bf n}$ points in the direction
perpendicular to the substrate surface, which we take at $y=0$. In this
case the SOC takes the form $\alpha \sigma^x q_z$. The spectrum of
excitations is comprised of two shifted parabolas, 
\begin{equation}\label{hsemi2}
{\epsilon}(q_z)={\hbar^2 q_z^2\over 2m_{\rm eff}}\pm\alpha q_z, 
\end{equation}
and is depicted in Fig.\ \ref{figss5}a. It is seen that SOC separates
the two spin projections but there is still an even number of Dirac
points in the right half of the Brillouin zone for any chemical potential. 
To change this one
must in addition apply a magnetic field $\bB$. If its direction is
perpendicular to $x$ then a gap opens up in the spectrum
(\ref{hsemi2}) at $q_z=0$ due to the Zeeman coupling, as illustrated
in Fig.\ \ref{figss5}b. Specifically,
for the field along $z$ the perturbation reads
$\delta{\cal H}=V_Z \sigma^z$
and the spectrum becomes
\begin{equation}\label{hsemi4}
{\epsilon}(q_z)={\hbar^2 q_z^2\over 2m_{\rm eff}}\pm\sqrt{\alpha^2 q_z^2+V_Z^2}, 
\end{equation}
When the chemical potential is tuned to lie inside the Zeeman gap, $|\mu|<V_z$, then the system exhibits a single Fermi point for $q_z>0$ and we expect it to enter the topological phase upon inducing the SC order. In the presence of a superconducting gap $\Delta$ the Kitaev criterion Eq.\ (\ref{maj1}) imposes 
\begin{equation}\label{hsemi5}
\sqrt{\mu^2+\Delta^2}<V_Z
\end{equation}
as a condition for the topological state. An explicit calculation once again confirms the existence of an unpaired Majorana mode at the end of the wire in this regime but we shall not reproduce it here. Also, we note that although SOC strength does not explicitly enter the criterion (\ref{hsemi5}) its presence is crucial for the emergence of the topological phase. 

As already mentioned this proposal has attracted by far the most attention from theorists and experimentalists alike and as a result much is known about this system beyond the simplest model outlined above. Among the key experimentally relevant issues are the effect of multiple bands (arising from the fact that real wires are not truly one-dimensional systems) and disorder, which is generically present in any solid state system, as well as details of the proximity effect for specific materials. Existing literature devoted to these issues has been reviewed in recent articles \cite{Alicea2012, Beenakker2012, Tewari2013}. An interested reader is referred to the above mentioned review articles for more in-depth discussion. 

Despite various challenges, semiconductor nanowires proximity coupled to superconductors are currently farthest along in terms of experimental realization of MZMs. We shall give a brief review of the existing experimental studies in Sec.\ \ref{Sec:ProspectsObserv} and provide a critical discussion of the claims that MZMs have been already observed in these systems.

\subsubsection{Helical Spin Chains}
\label{sec:spinchains}
In the proposals discussed thus far spin-orbit coupling played an
essential role in giving rise to a normal state with spin degeneracy
removed. SOC is fundamentally a relativistic effect and this is the
main reason why the associated energy scales remain relatively small
in solid state systems. We close this subsection with a description of
a simple system that can host MZMs but does not rely on
SOC. Consider a chain of magnetic atoms, such as Gd, Cr or Fe,
deposited on an atomically flat SC substrate. Each such magnetic moment
is known to create a bound state with the energy inside the SC gap,
known as the Shiba state \cite{Shiba1968}.  The distance between the magnetic atoms is
chosen such that the bound-state wavefunctions have a significant
overlap $t$. It turns out that the magnetic moments $\bS_j$ of the
atoms in this situation have a tendency to order in a co-planar
spiral. The 1D electron system formed of the
hybridized Shiba states then can be in the topological phase with unpaired MZMs localized at its ends \cite{Choy2011,Martin2012,Nadj-Perge2013}.

The spiral state and its stability against fluctuations and interactions has been studied in recent theoretical works \cite{Vazifeh2013,Klinovaja2013,Braunecker2013,Pientka2013} with some encouraging results. The main finding is that for energetic reasons the spiral pitch $G$ selfconsistently adjusts to the changing chemical potential so that the system remains in the topological phase. In addition, this `self-organized' topological state appears stable against the effects of fluctuations and interactions.

Structures composed of single atoms such as the 1D chain discussed here can now be built fairly routinely using techniques of scanning tunneling microscopy \cite{Manoharan}. The first attempt to construct the present system has been recently reported with some encouraging results and will be discussed in Sec.\ IV.E.

\subsection{Systems in Two Dimensions}

An important property of the solid-state realizations of Majorana particles in two dimensions is their non-trivial exchange statistics. This property is thought to harbor a unique potential for future applications in quantum computation in which operations would be topologically protected against the effects of decoherence. The recent surge of interest in Majorana fermions owes much to these prospects and the aim of this subsection is to explain the physics behind the phenomenon of non-Abelian exchange statistics as realized in simple models of 2D topological superconductors.

\subsubsection{Non-Abelian Exchange Statistics: General Considerations}
\label{sec:NonAbelian}

As noted before a pair of spatially separated MZMs $\gamma_{j1}$, $\gamma_{j2}$ can be thought of as forming one ordinary Dirac fermion represented by creation and annihilation operators $c^\dagger_j$, $c_j$ defined in Eq.\ (\ref{can1}). Let us imagine that we have $2N$ such well-separated MZMs arranged in a 2D plane. We also assume that there are no other low-energy degrees of freedom in this region of space, i.e.\ our MZMs are protected by a minigap. It is easy to see that the ground state of this system exhibits a $2^N$-fold degeneracy\footnote{In reality the degeneracy is only $2^{N-1}$ because of the electron parity considerations. $\nu=(\sum_j n_j)\mod 2$ represents the electron parity which is conserved in a fully gapped, isolated superconductor.} arising from two possible occupancies $n_j=0,1$ of each of the $N$ ordinary fermion states. The corresponding degenerate Hilbert space is spanned by basis vectors  
\begin{equation}\label{abel1}
|\Phi_{\{n_j\}}\rangle =|n_1,n_2,\dots n_N\rangle,
\end{equation}
where $n_j$ denotes the eigenvalue of the corresponding number operator
\begin{equation}\label{abel2}
\hat{n}_j=c^\dagger_jc_j={1\over 2}(1+i\gamma_{j1}\gamma_{j2}).
\end{equation}
The state vector $|\Psi\rangle$ composed of an arbitrary linear combination of the basis vectors $|\Phi_{\{n_j\}}\rangle$ can be used to encode quantum information. 

This way of encoding quantum information has two important advantages compared to many other schemes. First, as can be seen from the definition of the number operator in Eq.\ (\ref{abel2}), the information in each quantum bit is stored {\em nonlocally}. In order to read the information one must either bring the constituent MZMs close together  (to test if the combined fermionic state is filled or empty by a local measurement) or else perform a coherent measurement at two distant spatial positions.  Since the environment presumably cannot perform a non-local measurement, the information stored in the quantum bit $\hat{n}_j$ is thought to be immune to the effects of decoherence. One caveat here is that if there exist uncontrolled low-energy excitations in the system then the environment can potentially flip the quantum bit (without reading it) by tunneling a fermion into the qubit -- this is possible even if only one Majorana can be accessed. The existence of the minigap is therefore crucial for preserving the quantum information stored in a pair of MZMs.

The second key feature of the setup described above is the ability to manipulate the quantum information stored in $|\Psi\rangle$ in a topologically protected fashion. As we will demonstrate below {\em braiding} of MZMs --  performing adiabatic exchanges of their positions -- implements certain unitary transformations on the state vector $|\Psi\rangle$. The corresponding braid group turns out to be non-Abelian, meaning that unitary transformations describing individual exchanges do not in general commute. A sequence of exchanges performed on a properly initialized quantum state $|\Psi_i\rangle$ followed by a readout of the final state $|\Psi_f\rangle$ thus constitutes a topologically protected quantum computation. Unfortunately, it is known that the braid group realized by exchanging MZMs is not sufficiently rich to perform an arbitrary unitary transformation on $|\Psi_i\rangle$ which would be needed to implement a universal quantum computer. To achieve the latter the braid group must be supplemented by some `unprotected' operations  making the system vulnerable to decoherence, although in theory much less so than a non-topological quantum computer \cite{Nayak2008}. 

Majorana zero modes with the above non-Abelian exchange statistics have been theoretically proposed to exist in a number of 2D systems. Historically the first was the so called Moore-Read Pfaffian state \cite{Moore1991} that many believe describes the fractional quantum Hall state observed at $\nu={5\over 2}$ filling. Another is the thin film of a spin-polarized $p_x+ip_y$ superconductor, which may be realized in Sr$_2$Ru$_3$, although definitive evidence for this pairing state is still lacking \cite{Kallin2012}. More recently Fu and Kane proposed that a 2D topological superconductor with the requisite properties could arise at an interface formed between a 3D topological insulator and a conventional $s$-wave superconductor \cite{Fu2008}. MZMs are expected to be localized in the cores of Abrikosov vortices in such a 2D superconductor. In the following we shall focus on this model because it is closest in the spirit to our previous discussions and also because it might be most amenable to various practical tests of non-Abelian exchange statistics.

\subsubsection{Vortices in Fu-Kane Model}
\label{sec:FuKane}
\begin{figure}[t]
\includegraphics[width = 8.5cm]{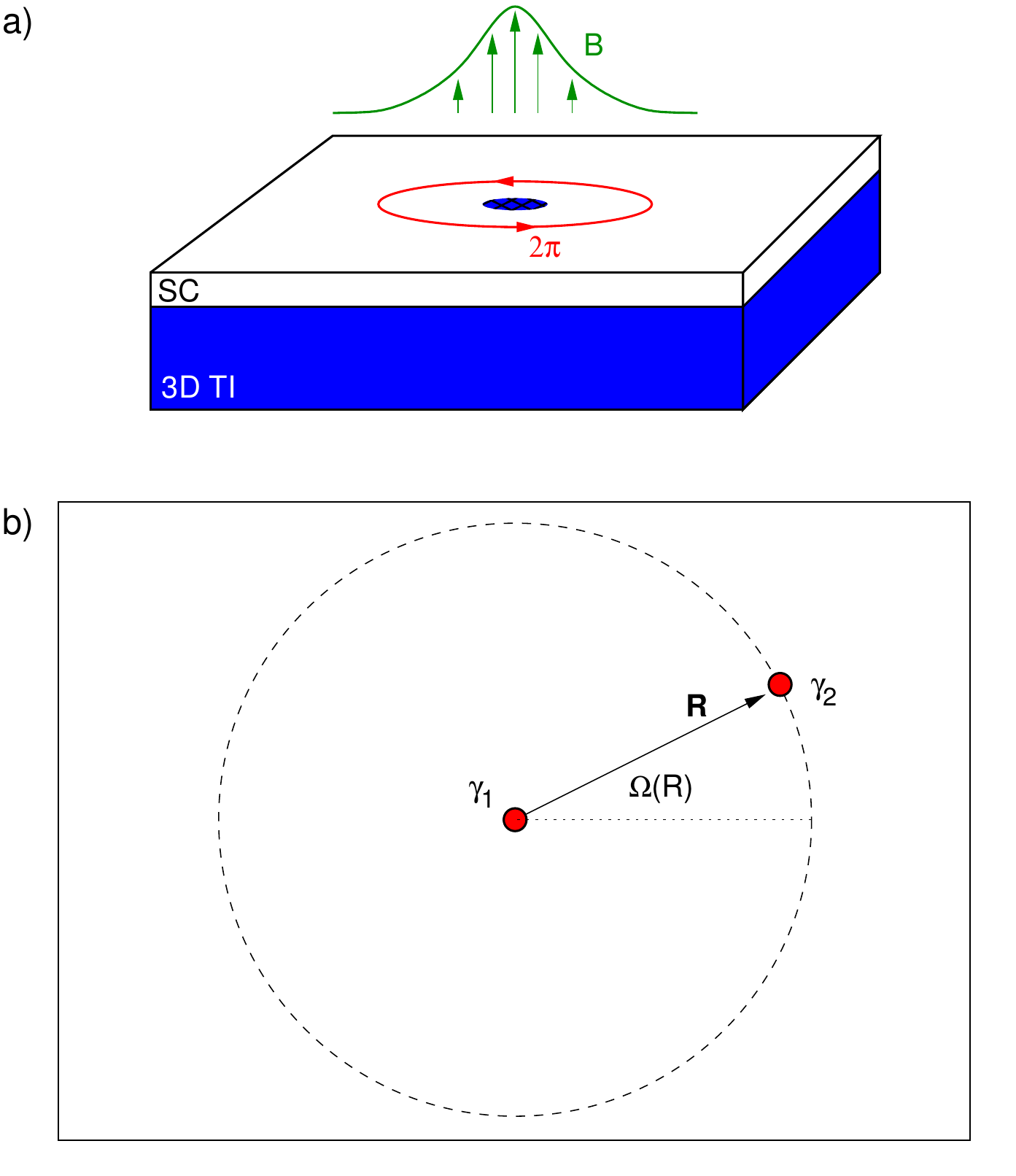}
\caption{ a) Schematic setup for the Fu-Kane model with a 3D TI
  covered with a thin layer of SC film. A vortex is depicted in the
  surface layer with a small core, phase winding $2\pi$ and a sketch
  of a magnetic field profile $\bB(\br)$. b) Vortex with Majorana
  zero mode $\gamma_2$ encircles a vortex placed at the origin with
  Majorana $\gamma_1$ in its core. 
}\label{figss7}
\end{figure}

Fu and Kane \cite{Fu2008} envisioned inducing superconductivity in the surface state
of a 3D topological insulator by covering it in with a thin film of an
ordinary $s$-wave superconductor such as Pb or Nb as depicted in Fig.\
\ref{figss7}a. Although more elaborate treatments are possible the simplest model that captures the essential physics of this situation consists of a Hamiltonian $h_0$ describing the TI surface defined in Eq.\ (\ref{h3dti1}) with the superconducting order included via the BdG formalism described in Sec.\ \ref{Sec:Theory}B. The resulting BdG Hamiltonian (\ref{hbdg1}) can be written in the following form
\begin{equation}\label{abel3}
H_{\rm BdG}(\br)= 
\begin{pmatrix}
0 & p_- & \Delta(\br) & 0 \\
p_+ & 0 & 0 & \Delta(\br) \\
\Delta^*(\br) & 0 & 0 & -p_- \\
0 & \Delta^*(\br) & -p_+ & 0
\end{pmatrix},
\end{equation}
where $p_{\pm}=p_x\pm ip_y$, and we have set $v=1$ and $\mu=0$ for the sake of simplicity. 

We are interested in finding the eigenstates of $H_{\rm BdG}(\br)$ in
the presence of an Abrikosov vortex. For our purposes the vortex is
defined as a point around which the phase of $\Delta(\br)$ winds by
$2\pi n$ with $n$ integer. More generally, an isolated Abrikosov
vortex also requires a magnetic flux $hc/2e$ spread in a flux tube
with a characteristic size $\lambda\simeq 10-10^3$\AA\  around the
vortex center (Fig.\ \ref{figss7}a). However, inclusion of the magnetic flux is unimportant for the existence of the MZM and we shall henceforth neglect it. For a vortex placed at the origin we thus have 
\begin{equation}\label{abel4}
\Delta(\br)=\Delta_0(r)e^{-i(n\varphi+\alpha)},
\end{equation}
where $\Delta_0(r)$ is a real function of the distance, $\varphi$ represents the polar angle and $\alpha$ denotes an arbitrary constant phase offset that will become important in our later discussion of vortex braiding. Single-valuedness of the Hamiltonian dictates that for $n$ nonzero  $\Delta_0(r)$ vanishes at the origin. Energy considerations further show that $\Delta_0(r)\sim r^{|n|}$ for small $r$.

To find the zero modes of  $H_{\rm BdG}(\br)$ in the presence of a vortex it is useful to first perform a unitary transformation $H_n=UH_{\rm BdG}U^{-1}$ with  
\begin{equation}\label{abel5}
U=
\begin{pmatrix}
1 & 0 & 0 & 0 \\
0 & 0 & 0 & 1 \\
0 & 0 & 1 & 0 \\
0 & 1 & 0 & 0
\end{pmatrix},
\end{equation}
which brings the Hamiltonian into the following off-diagonal form
\begin{equation}\label{abel6}
H_{n}= 
\begin{pmatrix}
0 & D_n \\
D^\dagger_n & 0 
\end{pmatrix}, \ \ \ 
D_{n}= 
\begin{pmatrix}
\Delta(\br) & p_- \\
-p_+ & \Delta^*(\br) 
\end{pmatrix}.
\end{equation}
The transformed Hamiltonian acts on the Nambu spinor $\hat\Psi_\br=(c_{\uparrow\br},c^\dagger_{\uparrow\br},-c^\dagger_{\downarrow\br},c_{\downarrow\br})^T$.
When looking for the zero modes the off-diagonal form of the Hamiltonian (\ref{abel6}) has a distinct advantage: the zero modes necessarily have the spinor structure $(\psi(\br),0)^T$ and $(0, \chi(\br))^T$ where $\psi(\br)$ and $\chi(\br)$ are two-component zero modes of $D^\dagger_n$ and $D_n$ respectively.
For a positive singly quantized vortex $(n=1)$ it is easy to show that there exists a normalizable zero mode of $D_1$ of the form
\begin{equation}\label{abel9}
\chi(\br)={1\over\sqrt{2}}
\begin{pmatrix}
e^{-i(\alpha/2-\pi/4)}\\
e^{i(\alpha/2-\pi/4)}
\end{pmatrix}
f_0(r), 
\end{equation}
with  $f_0(r)=Ae^{-\int_0^r\Delta_0(r')dr'}$,
while $D^\dagger_1$ does not have a normalizable zero mode. The field operator of the zero mode can be constructed following Eq.\ (\ref{psi2}) and reads
\begin{equation}\label{abel10}
\hat\psi_0={i\over\sqrt{2}}\int d^2r \left[e^{i(\alpha/2-\pi/4)}c_{\br\downarrow}-e^{-i(\alpha/2-\pi/4)}c^\dagger_{\br\downarrow}\right]f_0(r).
\end{equation}
As expected, the zero mode represents a Majorana particle, $\hat\psi_0^\dagger=\hat\psi_0$. We note that a singly quantized antivortex ($n=-1$) also possesses a zero mode, this time in the upper component of the spinor $\psi(\br)$ and is then composed of a spin-up electron and a spin-up hole.

\subsubsection{Vortex Exchange and Braiding}

\begin{figure}[t]
\includegraphics[width = 8.5cm]{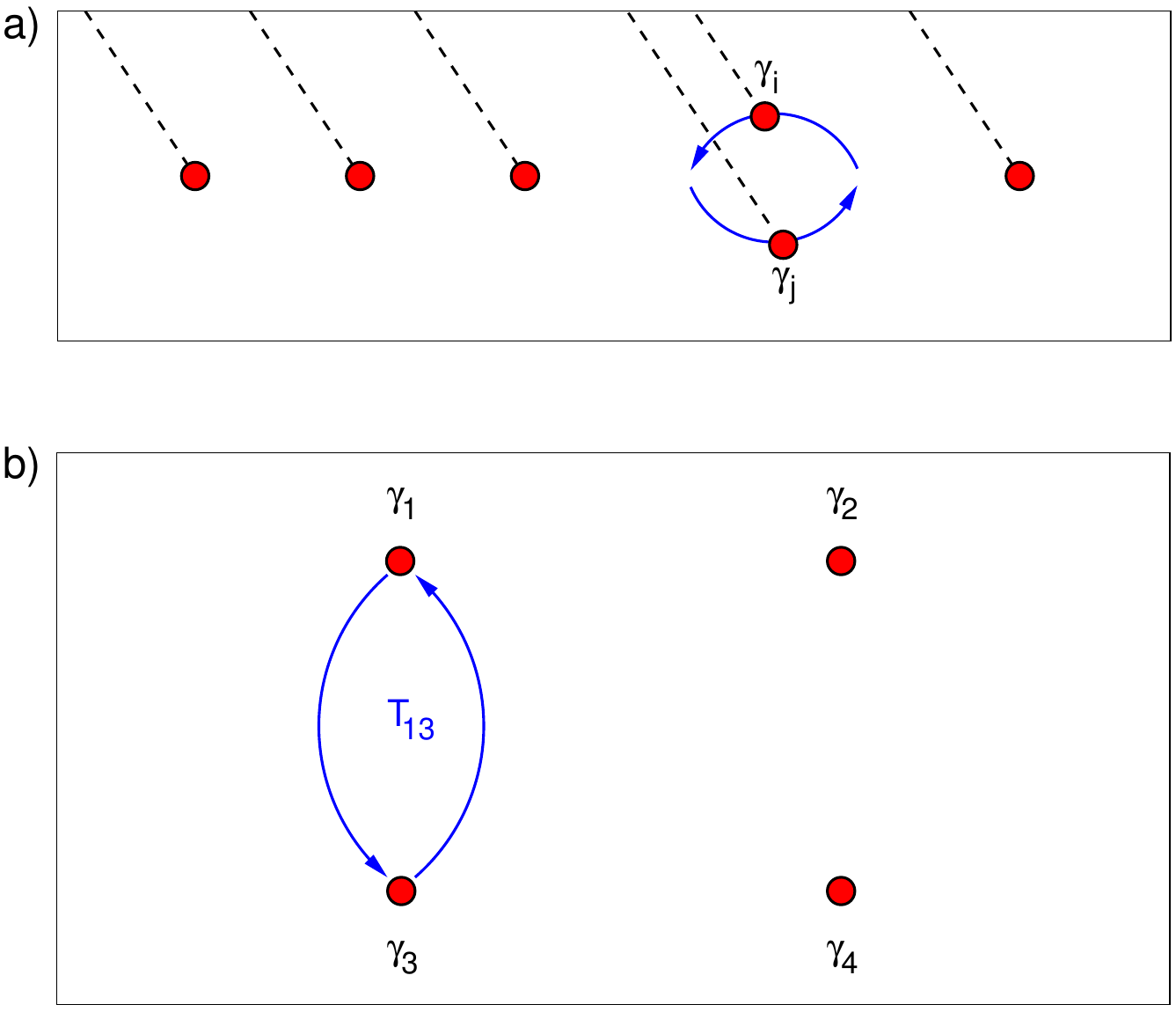}
\caption{ a) Upon exchange of two Majoranas 
  one of them must cross the branch cut implied by the wavefunction
  Eq.\ (\ref{abel12}). When the branch cuts are chosen as indicated by
  the dashed lines then counterclockwise exchange of $\gamma_i$ and
  $\gamma_j$ results in rules summarized in Eq.\ (\ref{abel14b}). 
b) Setup used for the simplest topologically
protected operations on the internal Hilbert space spanned by
four MZMs.
}
\label{figss8}
\end{figure}
Having established the existence of an MZMs in the core of a vortex in the Fu-Kane model we now proceed to discuss their statistics under exchange. These results follow from the earlier studies \cite{Moore1991,Read2000} but we follow the physically more transparent derivation given by Ivanov \cite{Ivanov2001}.  

We begin by considering the effect of a vortex encircling another vortex located at the origin. If their distance $d\gg\xi$ we can neglect the exponentially small splitting of the zero mode energies resulting from the wavefunction overlap and the only effect will be the change of SC phase near the origin due to the phase field produced by the distant vortex. In view of Eq.\ (\ref{abel4}) and by inspecting Fig.\ \ref{figss7}b this phase change  can be expressed as 
\begin{equation}\label{abel11}
\alpha(\bR)= \alpha_0+\Omega(\bR)+\pi
\end{equation}
where $\Omega(\bR)$ is the polar angle of the distant vortex and $\alpha_0$ denotes an arbitrary constant phase offset that we can adjust at will without affecting the physics. (The latter corresponds to the global U(1) phase of the condensate and does not have physical meaning.) Taking $\alpha_0=-{\pi\over 2}$ the wavefunction (\ref{abel9}) of the Majorana mode at the origin becomes  
\begin{equation}\label{abel12}
\chi_\bR(\br)={1\over\sqrt{2}}
\begin{pmatrix}
e^{-{i\over 2}\Omega(\bR)}\\
e^{{i\over 2}\Omega(\bR)}
\end{pmatrix}
f_0(r), 
\end{equation}
where the subscript $\bR$ reminds us that the wavefunction now depends on the position of the distant vortex.

As the distant vortex adiabatically encircles the origin counterclockwise over the time interval $t\in (0,T)$ the Majorana wavefunction $\chi(\br,t)$ acquires a Berry phase through the dependence of the instantaneous eigenstate $\chi_{\bR(t)}(\br)$ on the time parameter. The Berry phase reads
\begin{equation}\label{abel13}
\gamma({\cal C})=-{\rm Im}\oint_{\cal C}\langle\chi^{}_\bR|\nabla_\bR\chi^{}_\bR\rangle\cdot d\bR
-i\ln\left[\langle\chi_{\bR(0)}|\chi_{\bR(T)}\rangle\right],
\end{equation}
where the second term must be included in order to account for the fact that 
$\chi_\bR(\br)$ as defined in Eq.\ (\ref{abel12}) is not single valued as $\Omega\to \Omega+2\pi$. An explicit evaluation shows that the first term vanishes but the second term yields $\gamma({\cal C})=\pi$. This leads to the conclusion that upon being encircled by another singly quantized vortex the MZM wavefunction {\em changes its sign}. This is a direct consequence of the wavefunction being an equal superposition of an electron and a hole, which acquire a phase of $+\pi$ and $-\pi$ respectively upon the adiabatic change in the order parameter phase by $2\pi$.

The Majorana wavefunction of the distant vortex also changes sign because its local SC phase likewise advances by $2\pi$. If we denote the Majorana operators corresponding to the two vortices by $\gamma_1$ and $\gamma_2$ then the effect of the encircling operation can be encoded as
\begin{equation}\label{abel14}
\gamma_1\mapsto -\gamma_1, \ \ \ \gamma_2\mapsto -\gamma_2.
\end{equation}
For a set of $2N$ Majorana modes $\gamma_k$ an operation in which $\gamma_j$ encircles $\gamma_i$ can be implemented by a unitary transformation  
\begin{equation}\label{abel15}
\gamma_k\mapsto U_{ij}\gamma_kU^\dagger_{ij}, \ \ \ U_{ij}=\gamma_i\gamma_j.
\end{equation}

An adiabatic exchange of two Majoranas $\gamma_i$ and $\gamma_j$ can be thought of as a half of the encircling operation (two subsequent counterclockwise exchanges are equivalent to a single counterclockwise encircling operation). The unitary operator implementing such an exchange is therefore
\begin{equation}\label{abel15a}
T_{ij}=(U_{ij})^{1/2}={1\over\sqrt{2}}(1+\gamma_i\gamma_j). 
\end{equation}
Applying this unitary transformation we find the following rule governing such pairwise exchanges
\begin{equation}\label{abel14b}
\gamma_i\mapsto -\gamma_j, \ \ \ \gamma_j\mapsto \gamma_i, \ \ \ \gamma_k\mapsto \gamma_k,
\end{equation}
for $k\neq i,j$. These rules, first derived in this form by Ivanov \cite{Ivanov2001}, can be
intuitively understood by appealing to Fig.\ \ref{figss8}a. The form of $T_{ij}$ given in Eq.\ (\ref{abel15a}) also belies the non-Abelian structure of the braid group; it is easy to check that subsequent exchanges do not in general commute, e.g.\ $T_{12}T_{23}\neq T_{23}T_{12}$.

We close this subsection by working out a simple example that nicely
illustrates the action of the above derived transformations on
specific quantum states. Consider a system with 4 MZMs
$\gamma_1$, $\gamma_2$, $\gamma_3$, and $\gamma_4$ localized in
vortices and arranged as indicated in Fig.\ \ref{figss8}b. Of these we form two ordinary fermions 
\begin{equation}\label{abel15}
c_a={1\over 2}(\gamma_1+i\gamma_2), \ \ \ c_b={1\over 2}(\gamma_3+i\gamma_4),
\end{equation}
and label the resulting two-dimensional Hilbert space by the eigenvalues of the corresponding number operators $|n_a,n_b\rangle$. Let us consider the action of several operations on this state. Encircling $\gamma_3$ by $\gamma_1$ is implemented by $U_{31}=\gamma_3\gamma_1=(c^\dagger_a+c_a)(c^\dagger_b+c_b)$ and gives 
\begin{equation}\label{abel16}
|n_a,n_b\rangle \mapsto U_{31}|n_a,n_b\rangle=(-1)^{n_a}|\bar{n}_a,\bar{n}_b\rangle,
\end{equation}
where $\bar{n}=(1-n)\mod 2$ denotes a state with opposite occupancy to
$n$. The encircling operation performed between the constituent members of
different fermions thus reverses the occupancy of the states, e.g.\
$|0,0\rangle \mapsto |1,1\rangle$ or $|1,0\rangle \mapsto
-|0,1\rangle$. Meanwhile encircling between the Majorana members of
the same fermion just changes the overall sign of the state vector.

One can similarly work out the effect of exchanges, for instance 
\begin{eqnarray}\label{abel17}
 T_{12}|n_a,n_b\rangle &=& e^{i{\pi\over 4}(1-2n_a)}|n_a,n_b\rangle \\
 T_{31}|n_a,n_b\rangle &=&{1\over \sqrt{2}}\left( |n_a,n_b\rangle +(-1)^{n_a}|\bar{n}_a,\bar{n}_b\rangle\right), \nonumber 
\end{eqnarray}
We observe that exchanging two Majoranas belonging to the same fermion
$c$ merely attaches an (occupancy dependent) overall phase factor to
the state. Exchange of Majoranas belonging to different fermions,
however, creates a new entangled state. The
set of operations afforded by the braid group allows for non-trivial manipulations of the ground-state degenerate manifold but unfortunately 
does not permit creation of an arbitrary state
$|\Psi\rangle=\sum_{n_a,n_b}C_{n_a,n_b}|n_a,n_b\rangle$  from a given reference
state by repeated application of group elements $T_{ij}$.  As a result this
system cannot be used to perform a generic quantum computation. There
exist, however, theoretical proposals for solid state realizations of emergent
particles with more complicated non-Abelian statistics, e.g.\ the so
called ``Fibonacci anyons'', whose braid group is sufficiently rich to
permit construction of a universal quantum computer \cite{Nayak2008}.

\subsection{Experimental Observations}

\subsubsection{Quantum Wires and Other 1D Systems}

\begin{figure}[t]
\includegraphics[width = 8.0cm]{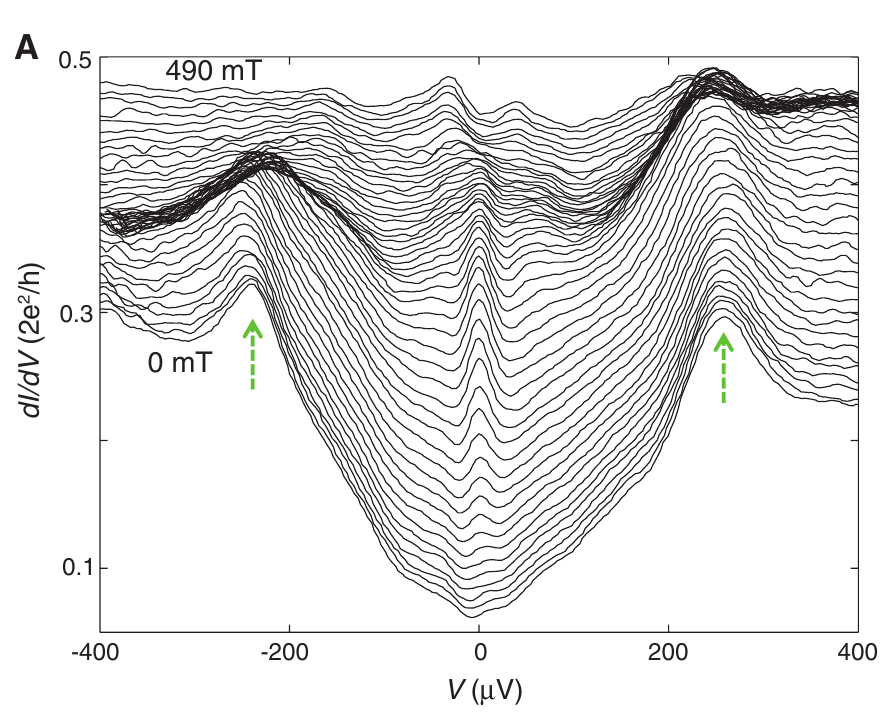}
\includegraphics[width = 8.0cm]{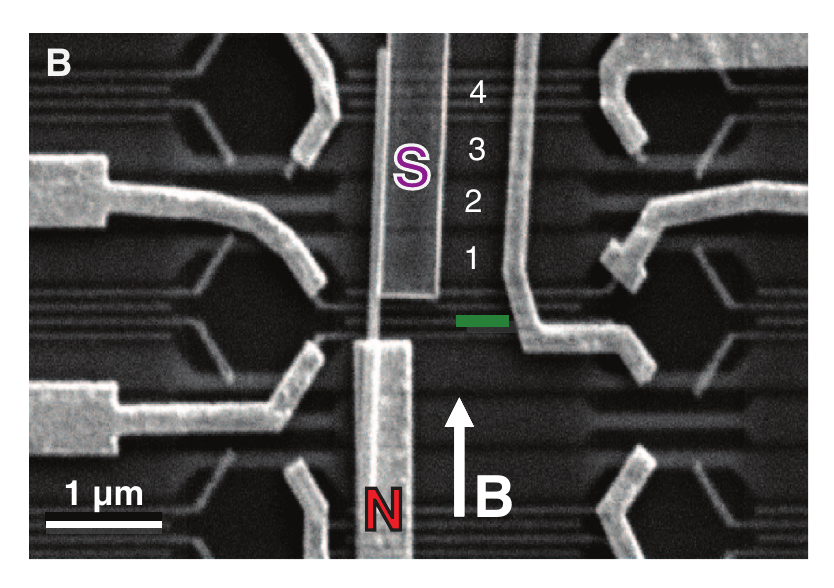}
\caption{The Delft experiment [figure adapted from \cite{Mourik2012}].  a) Tunneling conductance $g(V)$ as a function of the voltage bias showing a SC gap at low magnetic field and the emergence of a zero-bias peak attributed to the MZM at higher fields. b) Scanning electron microscope image of the device with normal (N) and superconducting (S) contacts attached to an InSb nanowire. 
}\label{figss9}
\end{figure}

By far the greatest progress to date in detecting the Majorana zero modes in solid state devices has been achieved in semiconductor quantum wires. Fig.\ \ref{figss9}a reproduces the original pioneering result of the Delft group \cite{Mourik2012} showing the historically first experimental evidence for the MZM. In the experiment a wire made of an InSb single crystal has been deposited on a substrate equipped with gates and contacted with superconducting and normal metal electrodes as depicted in Fig.\ \ref{figss9}b. According to the theory explained in Sec.\ \ref{sec:smwires} two Majoranas should appear at the ends of the SC sections of the InSb wire. Gates in the substrate are used to deplete the electron density in the section between the SC and normal metal electrodes and thus create a weak link. The existence of the MZM is then probed by measuring the tunneling current $I$ through this weak link under an applied voltage bias $V$. In this setup, to a very good approximation, differential tunneling conductance $g(V)=dI/dV$ is proportional to the density of states in the SC end of the wire adjacent to the tunneling contact. When the magnetic field $B$ is below about 90 mT the tunneling conductance shows a SC gap of about 260~$\mu$eV with no significant features at low energy (Fig.\ \ref{figss9}a). However, as the field is further increased a zero bias peak is seen to emerge in $g(V)$  which persists until about 400~mT and then disappears. This behavior is qualitatively consistent with the theoretical prediction for the MZM and has been interpreted as such.     

The result described above has been subsequently reproduced by several independent groups \cite{Deng2012,Das2012,Finck2013,Churchill2013} using variants of the setup indicated in Fig.\ \ref{figss9}b. In all cases a zero-bias peak has been reported at non-zero magnetic field consistent with the existence of MZM in this device. These experiments are now viewed as compelling evidence for Majorana particles in quantum wires although it must be noted that zero-bias anomalies often occur in superconducting systems in situations when MZMs are not expected to be present. For instance presence of disorder and multiple bands can conspire to produce zero-bias peaks in semiconductor wires (with the correct magnetic field dependence) even when the system is in the  topologically trivial phase \cite{Liu2012}. It is now generally thought that a truly conclusive experiment showing Majorana particles will have to test one of their other unique properties in addition to the zero-bias conductance peak. There exist several proposals in the literature to achieve this. The effects that can be probed include the fractional Josephson effect \cite{Kitaev2001}, quantized conductance in the ballistic regime \cite{Law2009,Wimmer2011}, various tests of non-locality \cite{Nilsson2008,Fu2010,Burnell2013} and the non-Abelian exchange statistics \cite{Alicea_NatPhys}. Of these only the first on the list has been thus far tested \cite{Rohkinson2012} with a positive result. Unfortunately it has been subsequently pointed out  
that the fractional Josephson effect can actually arise under certain conditions even for Josephson junctions formed of ordinary superconductors with no MZMs \cite{Berg2012}. At the time of this writing it thus appears that although compelling experimental signatures consistent with MZMs in semiconductor quantum wires have been observed by multiple groups, further experiments will be necessary to obtain truly unambiguous evidence.  

Experiments searching for MZMs in quantum wires made of topological insulators are ongoing but have not yet succeeded in producing evidence. Thus far studies have established the existence of coherent surface states in Bi$_2$Te$_3$ nanowires \cite{Peng2010} and proximity-induced superconducting order has likewise been demonstrated \cite{Nitin2011}.  The key obstacle facing
the observation of MZMs appears to be the significant
bulk contribution to the electron conduction. 

MZMs have not yet been observed in structures based on edge states of 2D topological insulators discussed in Sec.\ \ref{sec:2DTI}. This is chiefly because of the relative paucity of suitable 2D TI materials as well as the notorious difficulty with the fabrication and manipulation of the prototype system, the HgTe quantum well \cite{Franz2013}. This proposed realization awaits discovery of new 2D TI materials that are straightforward to grow and fabricate into suitable devices. A worldwide effort to achieve this goal is currently underway \cite{Liu2008,Knez2011,Weeks2011,Ezawa2012,Lindner2011}.  An innovative proposal has been recently formulated for a reliable transport detection of MZMs in the edge states of 2D TIs \cite{Beenakker2013} which could prove useful if the proposed setup were to be experimentally implemented. 

A  recent study \cite{Yazdani_Erice} has reported preliminary evidence for zero-bias peaks (obtained through scanning tunneling spectroscopy) associated with the ends of chains of magnetic Gd atoms deposited on the (110) surface of superconducting Pb crystals. These are consistent with MZMs discussed in Sec.\ \ref{sec:spinchains} and if confirmed could constitute an exciting new direction for Majorana research in solid state devices. An even more recent study by the same group \cite{Yazdani_Science} reported evidence for MZMs in the chains of Fe atoms on similar Pb surfaces, with Fe magnetic moments ordered ferromagnetically. This result can be understood provided one takes into account the strong SOC present in the Pb substrate \cite{Sau77}.

\subsubsection{2D Systems}

The Fu-Kane model discussed in Sec.\ \ref{sec:FuKane} and its variants  remain the focus of experimental studies in two dimensions. 
The experimental efforts thus far have focused on improving materials and devices with the goal of producing the correct conditions for the emergence of Majorana particles. Superconducting proximity effect has been achieved in surfaces of Bi$_2$Se$_3$ with Ti/Al electrodes \cite{Williams2012} and unconventional Josephson effect possibly indicative of Majorana physics has been reported in these devices. More recently proximity effect in the surface of Bi$_x$Sb$_{2-x}$Se$_3$ with chemical potential inside the bulk bandgap has been demonstrated \cite{Sungjae2013} paving the way for future detailed studies of MZMs. In addition, high-temperature superconductivity has been induced in Bi$_2$Se$_3$ flakes and films by interfacing with a cuprate superconductor  Bi$_2$Sr$_2$CaCu$_2$O$_{8+\delta}$ \cite{Burch2012,Wang2013}. Very recently, evidence for MZMs has been reported in the cores of magnetic vortices in a thin film of Bi$_2$Te$_3$ grown on the surface of a conventional superconductor NbSe$_2$ \cite{Xu2014}. Using scanning tunneling spectroscopy some features consistent with the theoretical prediction for the Fu-Kane model \cite{Chiu2011} have been  observed.


\section{Summary and Conclusions}
\label{Sec:conclusions}
The concept of the Majorana fermion, introduced in the seminal 1937 paper
\cite{Majorana1937}, remains more relevant today than at any previous
time. Although experimentally thus far unobserved in the realm of fundamental
particles, ongoing searches are now approaching the sensitivity required
to test the Majorana character of the leading candidate, the
neutrino. In addition, Majorana fermions are thought to play an
important role in resolving
some of the key outstanding questions in particle physics and
cosmology,  including leptogenesis (abundance in Universe of matter
over antimatter), nature and origin of the dark matter and the
relevance of supersymmetry to our understanding of elementary
particles. As discussed in Sec.\ \ref{Sec:MajFermNucPart}
supersymmetry requires the existence of superpartners that are
Majorana fermions (e.g.\ the photino, superpartner of the photon). If
such particles are stable they may well constitute the dark matter. Similarly,
one way to understand the abundance of matter over antimatter in our
Universe is through primordial processes involving the decay of heavy
Majorana neutrinos accompanied by CP violation in the lepton sector.

Majorana fermions also appear in the formal description of solids with 
superconducting order. In fact, as explained in Sec.\
II.B, quasiparticle excitations above the
ground state of any superconductor posses all the key attributes of
Majorana fermions: they are electrically neutral spin-${1\over 2}$
fermions indistinguishable from their antiparticles. Of great current
interest in solid state physics are Majorana zero modes, i.e.\
excitations that exist at zero energy in an otherwise gapped system
and are typically localized and
spatially separated from one another. In this situation one can
probe their Majorana character in a tabletop experiment and, due to
their unusual properties including exotic non-Abelian exchange
statistics, even exploit them to
perform a protected quantum computation. Compelling
experimental evidence exists for Majorana zero modes in semiconductor
quantum wires coupled to ordinary superconductors. Their properties
are currently a subject of intense experimental and theoretical
studies as are other solid state systems that have the potential to
harbor Majorana particles.    
 
Although extremely disparate in their physical manifestations the above
phenomena  stand unified by their description through the Majorana
equation. 
In more ways than one, Majorana's seminal work may thus contain clues to our
origins. By enabling new transformative technologies it may also pave
the way towards our future.

\section*{Acknowledgments}
This material is based upon work supported by the U.S. Department of Energy, Office of Science, Office of Nuclear Physics under Award Number LANL20135009 (SRE). We acknowledge support from NSERC and CIfAR (MF).
In addition, we are indebted to numerous colleagues for discussions
and correspondence that helped shape our understanding of the
subject. Of these, our special thanks go to J. Alicea,
C.W.J. Beenakker, A.M. Cook, C.-K. Chiu, T. Goldman, I. Herbut,
W. Louis, R. Mohapatra, N. Read, G. Refael, G. Senjanovi\'{c}, and M.M. Vazifeh.  
\bibliographystyle{apsrmp}

\end{document}